\documentclass{jfm}
\usepackage{graphicx}
\usepackage{epstopdf, epsfig}
\usepackage{amsmath}
\usepackage{mathrsfs}
\usepackage{booktabs}
\usepackage{subcaption}

\shorttitle{Compressible turbulent boundary layers over two-dimensional square-rib roughness}
\shortauthor{Y. T. Su, W. X. Huang and C. X. Xu}

\title{Compressible turbulent boundary layers over two-dimensional square-rib roughness}

\author{Youtian Su\aff{1},
  Wei-Xi Huang\aff{1}
 \and Chunxiao Xu\aff{1}\corresp{\email{xucx@tsinghua.edu.cn}}}

\affiliation{\aff{1}AML, Department of Engineering Mechanics, Tsinghua University, Beijing 100084, PR China}

\begin{document}

\maketitle

\begin{abstract}
Direct numerical simulations are performed to investigate the combined effects of surface roughness and wall heat transfer on spatially developing compressible turbulent boundary layers at $Ma=2.5$. The roughness consists of transverse square bars with $\lambda_x/k=8$ and $k^+ \approx 35$, under adiabatic and wall-cooling ($T_w/T_r = 0.5$) conditions. Dynamically, the conventional zero-moment method fails to yield a consistent zero-plane displacement for the present cavity-type roughness. Instead, a fitting-based optimization procedure is proposed to determine the kinematic virtual origin, which successfully restores the logarithmic behavior. Based on this displacement, Griffin--Fu--Moin (GFM) transformation outperforms the classical van Driest transformation in recovering outer-layer similarity for the velocity defect. Thermodynamically, the physical disparity between momentum form drag and the absence of a corresponding heat transfer mechanism disrupts the classical Reynolds analogy. The effective turbulent Prandtl number ($Pr_e$) deviates severely from unity within the roughness sublayer, leading to the breakdown of the classical Generalized Reynolds Analogy (GRA). To address this, a modified rough-wall GRA (rGRA) is formulated by introducing an equivalent slip-plane or reference-point boundary conditions, which accurately reconstructs the temperature-velocity relationship by bypassing the near-wall thermal heterogeneity. Finally, the refined strong Reynolds analogy (RSRA) is shown to maintain predictive accuracy for fluctuation intensities in the outer layer despite near-wall modulation by roughness and cooling.
\end{abstract}


\section{Introduction}\label{sec:intro}

Compressible wall-bounded turbulent flows are ubiquitous in high-speed aerospace applications, where aerodynamic heating and skin friction are intrinsically governed by the kinematic and thermodynamic characteristics of the boundary layer. For smooth walls, a well-established theoretical framework exists to describe these mean velocity and temperature profiles. Dynamically, the classical van Driest (VD) transformation \citep{van1951turbulent} performs well under adiabatic conditions but fails in diabatic flows with severe wall heat flux. Subsequent models, such as the Volpiani transformation \citep{volpiani2020data} and GFM transformation \citep{griffin2021velocity}, successfully address this limitation. Notably, the GFM achieves exceptional accuracy in mapping compressible velocity profiles to the incompressible law of the wall by treating viscous and Reynolds shear stresses independently. Thermodynamically, building upon early classical analogies \citep{busemann1931handbuch, crocco1932sulla, walz1962compressible}, \cite{zhang2014generalized} proposed the Generalized Reynolds Analogy (GRA). By introducing a generalized recovery factor, the GRA extends the rigorous mapping between mean velocity and temperature fields to highly non-adiabatic smooth-wall compressible turbulent boundary layers.

However, despite the profound success of these transformations and the GRA over ideal smooth surfaces, realistic aerospace vehicles are inevitably rough due to manufacturing tolerances, thermal ablation, or specific thermal protection designs. The introduction of surface roughness fundamentally alters near-wall coherent structures and the transport mechanisms of momentum and heat. Thus, smooth-wall-based aerodynamic and aerothermal models face severe challenges when applied to compressible rough-wall turbulence.

In the context of incompressible flows, the dynamic effects of surface roughness are well established. The most direct impact manifests in the near-wall region where roughness elements disrupt the viscous sublayer and, upon appropriate zero-plane displacement correction, induce a uniform downward shift in the logarithmic velocity profile (i.e., the roughness function $\Delta U^+$) \citep{clauser1954turbulent, hama1954boundary}. This velocity deficit can also be characterized by the equivalent sand-grain roughness $k_s$ \citep{nikuradse1933stromungsgesetze, schlichting1937experimental}. Building upon this near-wall behavior, the theoretical cornerstone of rough-wall turbulence research is Townsend's outer-layer similarity hypothesis \citep{townsend1976structure, raupach1991rough}. This hypothesis postulates that, given sufficient scale separation, the outer-layer flow remains independent of the specific surface topography and is consistent with smooth-wall behavior. However, unlike three-dimensional roughness, two-dimensional (2D) roughness (e.g., transverse square ribs or wavy walls) induces intense flow separation, free shear layers, and complex wake interference. Consequently, satisfying outer-layer similarity for 2D roughness typically demands far more stringent scale separation conditions \citep{krogstad2012turbulence, flack2014roughness}.

When compressibility is introduced, recovering and verifying this outer-layer similarity heavily relies on the aforementioned velocity transformations. For instance, \cite{latin2000flow} investigated an adiabatic boundary layer over d-type (pitch-to-height ratio $\lambda_x/k=4$) square-ribs roughness at $Ma=2.9$, demonstrating that the VD transformation successfully recovers the outer-layer similarity observed in incompressible flows. More recently, \cite{modesti2022direct} examined a cold-wall channel with cube roughness at $Ma=0.3 \sim 4$, and \cite{cogo2025surface} studied an adiabatic boundary layer over cube roughness at $Ma=0.3$ and $2.0$. They systematically evaluated the performance of the VD, Volpiani, GFM, and Hasan \citep{hasan2023incorporating} transformations. While all transformations exhibit good Mach-number independence for adiabatic cases, discrepancies emerge in the presence of wall heat flux. Notably, the latter three transformations (Volpiani, GFM, Hasan) outperform the VD transformation in preserving outer-layer similarity.

Furthermore, when applying velocity transformations to such complex boundary layers, accurately determining the virtual origin, or zero-plane displacement, is a prerequisite for recovering the logarithmic law. In incompressible flows, although the zero-moment height method introduced by \cite{jackson1981displacement} is widely adopted, it is not universally applicable to all surface conditions \citep{chan2015systematic, macdonald2018direct}. For compressible roughness, \cite{modesti2022direct} resorted to an empirical value of $0.9k$ for their zero-plane displacement. In contrast, for compressible rough-wall cases featuring wavy or  three-dimensional sinusoidal topographies \citep{tyson2013numerical, aghaei2023supersonic, chen2024energy}, the determination of the zero-plane displacement is typically unnecessary since the mean surface elevation can be directly adopted as the virtual origin.

Compared to momentum transport, the thermodynamic coupling mechanism in compressible rough-wall turbulence is considerably more complex. There is a fundamental distinction between heat and momentum transfer: while the pressure drag (form drag) induced by roughness elements profoundly alters momentum transport, there is no analogous heat transfer mechanism in the heat conduction process \citep{dipprey1963heat}. This severe physical asymmetry indicates that the similarity between momentum and heat transport is significantly disrupted over rough walls, particularly under strong wall heat flux conditions. In the literature, for a three-dimensional sinusoidal rough-wall case, \cite{wang2024roughness} found that the classical Walz's equation maintains good accuracy across the entire boundary layer profile under cold-wall conditions ($T_w/T_r = 0.84, 0.43$). However, Modesti \textit{et al.} \citep{modesti2022direct} observed in a cold-wall channel with cube roughness ($T_b/T_w = 0.5 \sim 0.994$) that while the GRA yields satisfactory predictions at $Ma=2$, the prediction deviation notably increases at $Ma=4$. Furthermore, analyzing adiabatic boundary layers over various 3D block roughness topologies at $Ma=2$, \cite{cogo2025surface, cogo2025development} pointed out that although the GRA provides reasonable predictions in the outer layer ($y/\delta > 0.5$), a certain degree of deviation persists within the roughness sublayer near the roughness elements.

Regarding the fluctuating flow field, the Strong Reynolds Analogy (SRA), first proposed by \cite{morkovin1962effects}, and its evolved models are widely employed to quantify the statistical relationship between temperature and velocity fluctuation intensities. While the classical SRA relies on the stringent assumptions of unity Prandtl number and zero total temperature fluctuations, subsequent models have made significant improvements. Gaviglio \citep{gaviglio1987reynolds} presented the SRA (GSRA) for ideal gas by introducing the concept of turbulent fluctuation length to link fluctuation intensities with mean gradients, overcoming the limitations of the classical assumptions. Building on this, Huang \textit{et al.} \citep{huang1995compressible} extended SRA (HSRA) by incorporating the turbulent Prandtl number ($Pr_t$), which effectively enhanced the model's applicability under various non-adiabatic wall conditions. Recently, Huang \textit{et al.} \citep{huang2025refined} conducted an in-depth analysis of the transport equations for velocity and temperature fluctuations. Based on the strong linear similarity mechanism between the time scales of their respective turbulent production terms, they proposed the Refined SRA (RSRA). This improvement leads to more accurate predictions of fluctuation intensities over smooth walls. \cite{cogo2025development} demonstrated that for adiabatic rough-wall cases, the classical SRA maintains good predictive accuracy, with its error distributions collapsing closely onto those of smooth-wall counterparts.

In summary, although existing dynamic velocity transformations and thermodynamic analogies have achieved notable progress over smooth and mildly rough surfaces, it remains to be further investigated whether momentum and heat transport over walls with strong roughness effect can still be mapped back to incompressible or smooth-wall behaviors via transformations and analogies. Dynamically, under classical compressible smooth-wall transformations, the recovery of the logarithmic law, the evolution of the roughness function, and the maintenance of outer-layer similarity are strongly influenced by 2D fully rough topographies compared to 3D or milder roughness counterparts. Thermodynamically, while previous studies indicate that the classical GRA exhibits a certain degree of predictive agreement in specific rough-wall cases, its predictive accuracy typically exhibits a systematic degradation, with the deviation amplifying both as the Mach number increases and as the distance to the wall decreases. This phenomenon largely reflects the fundamental physical disparity between drag and heat transfer mechanisms. These observations indicate that under boundary conditions superimposing intensified roughness effects with severe wall heat flux, near-wall perturbations will violate the assumptions of wall-normal conserved quantities inherent in smooth-wall theories (e.g., the effective turbulent Prandtl number $Pr_e \approx 1$ in the GRA). Consequently, the applicability of both the classical GRA predictive model and the SRA fluctuating framework is challenged.

The paper is organized as follows. Section \ref{sec:Methodology} introduces the computational methodology, detailing the dual-domain flow configuration, numerical methods, and the double-averaging statistical framework. Section \ref{sec:results_discussion} presents a systematic analysis of the simulation results, beginning with the overall flow characterization (\S\ref{sec:flow_characterization}) and drag distribution (\S\ref{sec:Drag_distribution}). Section \ref{sec:Mean_velocity_profiles} proposes a fitting-based optimization procedure to determine the physically consistent zero-plane displacement, followed by evaluations of the mean velocity profiles under various dynamic transformations. Section \ref{sec:Velocity_fluctuations} investigates the behavior of turbulent velocity fluctuations. In Section \ref{sec:GRA}, we introduce a modified rough-wall generalized Reynolds analogy (rGRA) to better describe the mean temperature-velocity relationship under the combined effects of surface roughness and wall heat transfer. Section \ref{sec:SRA} critically assesses the predictive accuracy of several strong Reynolds analogy (SRA) variants. Finally, the main conclusions are summarized in Section \ref{sec:Conclusions}.

\section{Methodology}\label{sec:Methodology}

\subsection{Flow configuration and governing equations}

\begin{figure}
    \centering

    \includegraphics[width=0.85\textwidth]{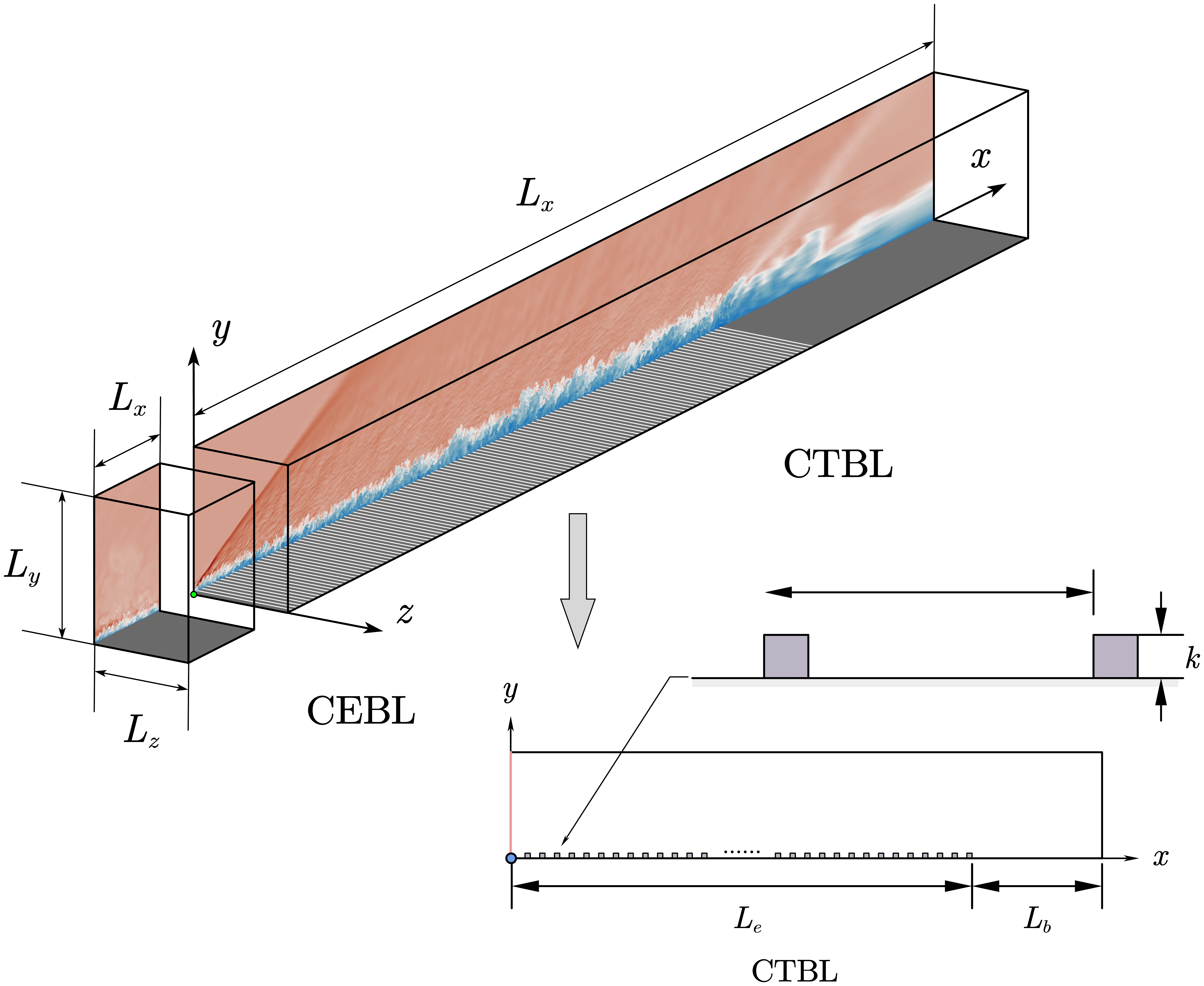}

    \caption{Schematic of the computational domain configurations, comprising the upstream compressible equivalent boundary layer (CEBL) domain and the downstream main compressible turbulent boundary layer (CTBL) domain. The green dot indicates the origin of the coordinate system, with the streamwise direction aligned with the positive $x$-axis. The inset in the bottom right corner shows the $xy$-plane cross-section of the CTBL domain.}
    \label{fig:computational_setup}
\end{figure}

The computational framework utilizes a dual-domain strategy consisting of two coupled domains, as schematically illustrated in figure \ref{fig:computational_setup}: an auxiliary domain simulating a compressible equivalent boundary layer (CEBL) to generate realistic inflow turbulence \citep{li2024inflow}, and a main domain for the spatially developing compressible turbulent boundary layer (CTBL). Four representative cases are investigated in the main domain, encompassing combinations of smooth and rough walls under different thermal conditions (Table \ref{tab:flow_parameters_split}). 

The flow is assumed to be a calorically perfect gas with a constant ratio of specific heats $\gamma = 1.4$. The freestream Mach number is set to $M_\infty = 2.5$, and the freestream temperature is 270 K. The fluid properties are governed by the ideal gas equation of state, $p = \rho R T$. The dynamic viscosity $\mu$ is evaluated using Sutherland's law, and the Prandtl number is strictly kept constant at $Pr = 0.71$. The wall-to-recovery temperature ratio is prescribed as $T_w/T_r = 1.0$ for the adiabatic cases and $T_w/T_r = 0.5$ for the cold-wall cases.

In the main CTBL domain, the flow is governed by the standard three-dimensional, compressible Navier--Stokes equations. The conservation of mass, momentum, and energy are respectively expressed as
\begin{equation}
    \frac{\partial \rho}{\partial t} + \frac{\partial \rho u_j}{\partial x_j} = 0,
\end{equation}
\begin{equation}
    \frac{\partial \rho u_i}{\partial t} + \frac{\partial \rho u_i u_j}{\partial x_j} = -\frac{\partial p}{\partial x_i} + \frac{\partial \sigma_{ij}}{\partial x_j},
\end{equation}
\begin{equation}
    \frac{\partial \rho E}{\partial t} + \frac{\partial \rho u_j H}{\partial x_j} = \frac{\partial (\sigma_{ij}u_i - q_j)}{\partial x_j},
\end{equation}
where $\rho$ is the fluid density, $u_i$ is the velocity vector component, $p$ is the pressure, $E$ and $H$ represent the total energy and total enthalpy, respectively; $\sigma_{ij}$ is the viscous stress tensor, and $q_j$ is the heat flux vector based on Fourier's law. 

To provide a high-fidelity, fully developed turbulent inflow for the main domain, the compressible CEBL method proposed by \cite{li2024inflow} is employed in the auxiliary domain (Figure \ref{fig:computational_setup}). The target friction Reynolds number for the generated smooth-wall inflow is $Re_\tau = 300$. Unlike traditional recycling-rescaling methods, the CEBL approach operates under a streamwise-equilibrious assumption by applying periodic boundary conditions in the streamwise direction, which drastically reduces the unphysical adjustment length. To analytically recover the mean mass, momentum, and energy balances of a spatially developing boundary layer at the designated Reynolds number, wall-normal distributions specific source terms ($M$, $F$, and $Q$) are intrinsically introduced into the governing equations. Consequently, the governing equations solved in the auxiliary CEBL domain are written as
\begin{equation}
    \frac{\partial \rho}{\partial t} + \frac{\partial \rho u_j}{\partial x_j} = -M,
\end{equation}
\begin{equation}
    \frac{\partial \rho u_i}{\partial t} + \frac{\partial \rho u_i u_j}{\partial x_j} + \frac{\partial p}{\partial x_i} - \frac{\partial \sigma_{ij}}{\partial x_j} = -F \delta_{i1},
\end{equation}
\begin{equation}
    \frac{\partial \rho E}{\partial t} + \frac{\partial \rho H u_j}{\partial x_j} - \frac{\partial (\sigma_{ij}u_i - q_j)}{\partial x_j} = -Q.
\end{equation}

Periodic boundary conditions are imposed in the streamwise direction of the CEBL domain. The turbulent fluctuations generated within this auxiliary domain are extracted and fed directly into the inlet boundary of the main CTBL domain. The bottom surface is treated as a no-slip solid wall, while periodic boundary conditions are enforced in the spanwise direction. The top boundary employs a non-reflecting boundary condition to prevent the unphysical reflection of acoustic waves back into the flow field. Additionally, at the downstream exit of the main CTBL domain, a supersonic outflow boundary condition is prescribed.

The geometric parameters of the surface roughness are determined based on the inner scales of the baseline smooth-wall boundary layers (M2p5SA and M2p5SC). The roughness elements are modeled as two-dimensional spanwise square bars. To achieve a physically consistent perturbation, the roughness height $k$ is set to $y^+ = 24.3$, based on the \textit{a posteriori} statistics of the baseline smooth-wall cases at the target friction Reynolds number of $Re_\tau \approx 510$. This criterion yields a uniform roughness height of $k = 0.086\delta_{in}$ for the adiabatic rough-wall case (M2p5RA) and $k = 0.080\delta_{in}$ for the cold rough-wall case (M2p5RC), where $\delta_{in}$ is the boundary layer thickness generated by the CEBL, which serves as the inflow condition for the CTBL. The streamwise pitch of the roughness elements is fixed at $\lambda_x = 8k$, which is consistent with the canonical configurations in previous studies \citep{lee2007direct,lee2011direct}. For both rough-wall cases, the effective physical domain accommodates 116 roughness periods.

For the auxiliary CEBL domain, the streamwise length is kept constant at $L_{x} = 10\delta_{in}$, with the wall-normal and spanwise dimensions matching those of the corresponding main CTBL domain. The dimensions of the main CTBL domains are summarized in table \ref{tab:computational_details}. The streamwise length of the main domain consists of an effective region ($L_e$) and a buffer zone ($L_b$). A buffer zone of length $L_b = 30\delta_{in}$ is applied to all cases to smoothly damp the outgoing flow and prevent spurious numerical reflections. Notably, the bottom wall within this buffer zone is kept smooth for all four cases.

The spanwise width $L_z$ is expanded to $9.0\delta_{in}$ for the rough-wall cases, compared to $5.9\delta_{in}$ for the smooth-wall cases. This enlargement is necessary to accommodate the more energetic and larger-scale turbulent structures that develop within the thicker rough-wall boundary layer, thereby preventing artificial periodicity in the spanwise direction.

\begin{table}
    \begin{center}
    \def~{\hphantom{0}}
    \setlength{\tabcolsep}{2.1pt} 
    \begin{tabular}{lcccc c lcccc}

        \multicolumn{5}{c}{Present Simulations} && \multicolumn{5}{c}{Reference Literature} \\[3pt]

        Case & Surf. & $M_\infty$ & $T_w/T_r$ & $Re_{\tau}$ &\, \, \, \, \,& Case & Surf. & $M_\infty$ & $T_w/T_r$ & $Re_{\tau}$ \\[3pt]

        M2p5SA & S & 2.5 & 1.0 & ~510 && \cite{zhang2018direct}         & S & 2.5  & 1.0 & ~510 \\
        M2p5SC & S & 2.5 & 0.5 & ~510 && \cite{volpiani2018effects}     & S & 2.28 & 0.5 & ~511 \\
        M2p5RA & R & 2.5 & 1.0 & 1050 && \cite{pirozzoli2011turbulence} & S & 2.0  & 1.0 & 1000 \\
        M2p5RC & R & 2.5 & 0.5 & 1050 && \cite{schlatter2010assessment} & S & 0   & --  & 1047 \\
               &   &     &     &      && \cite{eitel2014simulation}     & S & 0   & --  & 2478 \\

    \end{tabular}
    \caption{Physical parameters and flow conditions of the present simulations (left) alongside the reference literature selected for subsequent validation and analysis (right). In the nomenclature of the present cases, `S' and `R' denote smooth and rough surfaces (as also abbreviated in the Surf. column), while `A' and `C' indicate adiabatic ($T_w/T_r = 1.0$) and cold ($T_w/T_r = 0.5$) wall conditions, respectively. The parameter $Re_{\tau}$ represents the target friction Reynolds number of the boundary layer for the present simulations.}
    \label{tab:flow_parameters_split}
    \end{center}
\end{table}

\subsection{Numerical methods and computational grid}

The simulations of the turbulent boundary layers are performed using the high-fidelity finite-difference solver OpenCFD \citep{li2010direct}. The governing equations are integrated in time using an explicit 3rd-order Runge-Kutta scheme. For the spatial discretization of the convective terms, distinct numerical strategies are employed in the two computational domains. In the auxiliary CEBL domain, the convective fluxes are evaluated using a standard 7th-order WENO scheme. In the main CTBL domain, a 7th-order hybrid scheme is adopted to minimize numerical dissipation while preserving shock-capturing capabilities. This hybrid approach dynamically blends a low-dissipation 7th-order upwind scheme with 7th- and 5th-order WENO schemes. The viscous fluxes are computed using a 6th-order central difference scheme. 

To represent the rough-wall geometry, a sharp-interface ghost-cell immersed boundary method (IBM) is implemented \citep{de2020sharp}. In the immediate vicinity of the immersed solid boundaries, a stencil reduction method is applied to the spatial finite-difference operators. This specific treatment ensures that the computational stencil incorporates at most one immersed solid point, thereby maintaining numerical stability and boundary accuracy near the complex fluid-solid interfaces.

The computational domain is discretized using a structured Cartesian mesh. For the main CTBL domain, the global grid dimensions are $N_x \times N_y \times N_z = 2270 \times 300 \times 310$ for the smooth-wall cases and $6980 \times 300 \times 473$ for the rough-wall cases. The auxiliary CEBL domains utilize $N_x = 338$ grid points in the streamwise direction, while sharing identical wall-normal and spanwise grid distributions with their respective main domains. To adequately resolve the flow physics near the complex geometries, local grid refinement is applied in the immediate vicinity of the roughness elements. Specifically, a single spanwise square bar is resolved by $N_x \times N_y = 21 \times 37$ grid points for the adiabatic rough-wall case (M2p5RA) and $21 \times 35$ grid points for the cold rough-wall case (M2p5RC).

In the wall-normal direction, a custom grid stretching strategy is designed and applied across all simulation cases. The mesh is densely clustered near the wall to resolve the viscous sublayer and gradually stretched outwards. To adequately capture the energetic turbulent structures within the significantly thicker rough-wall boundary layers, the maximum grid spacing within the boundary layer is strictly maintained up to $y = 4.9\delta_{in}$ before undergoing further stretching towards the freestream. Additionally, a uniform grid spacing is adopted in the spanwise direction, and this grid size is kept identical across all simulation cases.

Target stations for statistical analysis are selected at $Re_\tau = 510$ for the smooth-wall cases and $Re_\tau = 1050$ for the rough-wall cases. The detailed grid allocations and the local inner-scaled resolutions at these target stations are summarized in table \ref{tab:computational_details}, confirming the direct numerical simulation (DNS) fidelity of the current database. Unless otherwise specified, the subsequent analyses regarding the mean drag, flow profiles and turbulence statistics are performed at these specific locations.

\begin{table}
    \begin{center}
    \def~{\hphantom{0}}
    \setlength{\tabcolsep}{2.5pt}
    \begin{tabular}{lcccccccc}

        Case & $(L_x \times L_y \times L_z)/\delta_{in}^3$ & $N_x \times N_y \times N_z$ & $k/\delta_{in}$ &  $x_t/\delta_{in}$ & $\delta/\delta_{in}$  & $\Delta x^+$ & $\Delta y^+$ & $\Delta z^+$ \\[3pt]

        M2p5SA & $95 \times 12.5 \times 5.9$  & $2270 \times 300 \times 310$ & --     & 59.0 & 1.84 & 8.21 & 0.602--9.45 & 5.28 \\
        M2p5SC & $95 \times 12.5 \times 5.9$  & $2270 \times 300 \times 310$ & --     & 40.2 & 1.71 & 8.82 & 0.647--10.1 & 5.67 \\
        M2p5RA & $109 \times 12.5 \times 9.0$ & $6980 \times 300 \times 473$ & 0.086  & 48.5 & 2.49 & --   & --           & --   \\
        M2p5RC & $103 \times 12.5 \times 9.0$ & $6980 \times 300 \times 473$ & 0.080  & 40.2 & 2.41 & --   & --           & --   \\

    \end{tabular}
    \caption{Computational domain, grid dimensions, and \textit{a posteriori} resolution details for the present simulation cases. $L_{x,y,z}$ and $N_{x,y,z}$ denote the domain lengths and the number of grid points in the streamwise, wall-normal, and spanwise directions, respectively. The parameter $x_t$ indicates the streamwise location where the target friction Reynolds number ($Re_{\tau}$ reported in Table \ref{tab:flow_parameters_split}) is achieved, and $\delta$ represents the local boundary layer thickness at this specific location.}
    \label{tab:computational_details}
    \end{center}
\end{table}

\subsection{Statistical framework and double-averaging method}

To quantify the inherently three-dimensional and heterogeneous flow structures induced by the surface roughness, a double-averaging method is employed. In the context of compressible turbulent flows, both Reynolds and Favre averaging procedures are required to formulate the statistical framework.

For an instantaneous flow quantity $\phi(x, y, z, t)$, the standard Reynolds decomposition is defined as
\begin{equation}
    \phi(x, y, z, t) = \overline{\phi}(x, y) + \phi'(x, y, z, t),
\end{equation}
where $\overline{\phi}(x, y)$ represents the Reynolds-averaged quantity, which is further averaged in the homogeneous spanwise direction due to the geometric configuration, and $\phi'$ denotes the associated turbulent fluctuation. For compressible flows, the Favre averaging is introduced:
\begin{equation}
    \phi(x, y, z, t) = \tilde{\phi}(x, y) + \phi''(x, y, z, t),
\end{equation}
where $\tilde{\phi} = \overline{\rho \phi} / \overline{\rho}$ is the Favre mean, and $\phi''$ is the Favre turbulent fluctuation. 

In the presence of rough walls, this mean flow remains spatially heterogeneous in the streamwise direction due to the stationary wakes behind individual roughness elements. To isolate these stationary spatial variations from the background turbulence, an intrinsic spatial averaging operator is introduced. For the analysis of mean profiles in the subsequent sections, the spatial average of a physical quantity is evaluated by integrating over a roughness period at a constant $y$ plane. To distinguish the fluid phase from the solid roughness elements, a phase indicator function $\Gamma(x, y)$ is introduced, which is independent of the spanwise coordinate $z$ owing to the two-dimensional geometry of the roughness. It is defined as $\Gamma = 1$ in the fluid phase and $\Gamma = 0$ in the solid phase. The intrinsic spatial average of a time-averaged quantity $\overline{\phi}$ is thus mathematically defined as
\begin{equation}
    \langle \overline{\phi }\rangle (x_n+\frac{\lambda _x}{2},y)=\frac{\int_{x_n}^{x_n+\lambda _x}{\overline{\phi }(x,y)\Gamma (x,y)dx}}{\int_{x_n}^{x_n+\lambda _x}{\Gamma (x,y)dx}},
    \label{eq:spatial_average}
\end{equation}
where the integration is performed over a streamwise spatial period $\lambda_x$, starting from the $n$-th roughness unit located at $x_n$. 

Consequently, the double-averaging (triple decomposition) is expressed as
\begin{equation}
    \phi(x, y, z, t) = \langle \overline{\phi} \rangle(y) + \overline{\phi}^d(x, y) + \phi'(x, y, z, t).
\end{equation}
Here, $\overline{\phi}^d(x, y) = \overline{\phi}(x, y) - \langle \overline{\phi} \rangle(y)$ represents the spatial fluctuation (or dispersive component), which characterizes the time-independent spatial variations induced by the 2D roughness elements. All statistical results presented in the subsequent sections are obtained by averaging over sufficient flow-through times to ensure statistical convergence. 

For the sake of brevity in the subsequent discussion, unless explicitly specified otherwise, all wall-normal profiles of physical quantities are inherently spatially averaged over the corresponding roughness periods. Consequently, the spatial averaging operator $\langle \cdot \rangle$ is hereafter omitted from the notations, and the variables presented in the profiles (e.g., $\phi$ or $\overline{\phi}$) should be interpreted as their spatially averaged counterparts $\langle \phi \rangle (y)$ or $\langle \overline{\phi} \rangle (y)$.

\section{Results and discussion}\label{sec:results_discussion}

\subsection{Overall flow characterization} \label{sec:flow_characterization}

\begin{figure}
    \centering
    \begin{subfigure}{\textwidth}
        \centering
        \setlength{\unitlength}{\textwidth} 
        \begin{picture}(1, 0.25) 
            \put(0,0){\includegraphics[width=\textwidth]{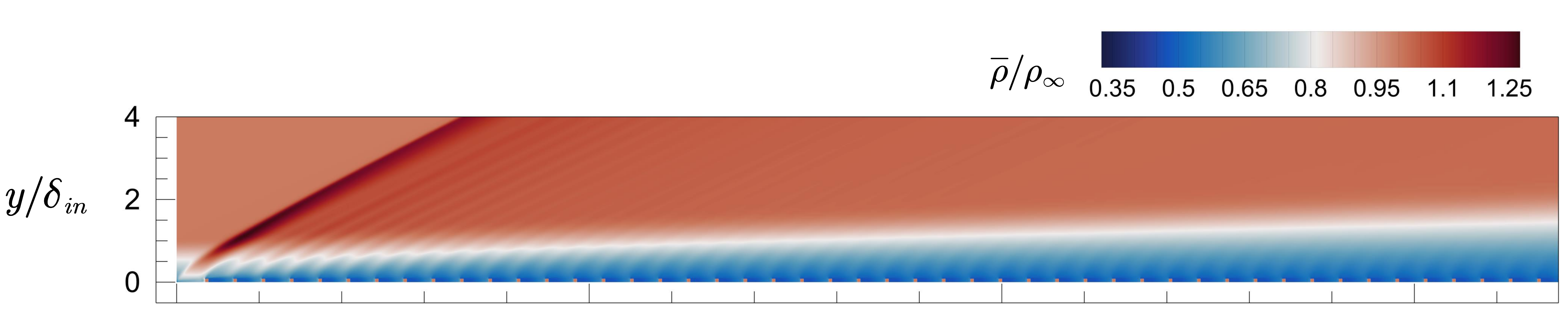}}
            \put(0.01, 0.16){\textbf{(\textit{a})}} 
        \end{picture}
        \label{fig:density_adiabatic}
    \end{subfigure}

    \vspace{-0.4cm} 

    \begin{subfigure}{\textwidth}
        \centering
        \setlength{\unitlength}{\textwidth}
        \begin{picture}(1, 0.25) 
            \put(0,0){\includegraphics[width=\textwidth]{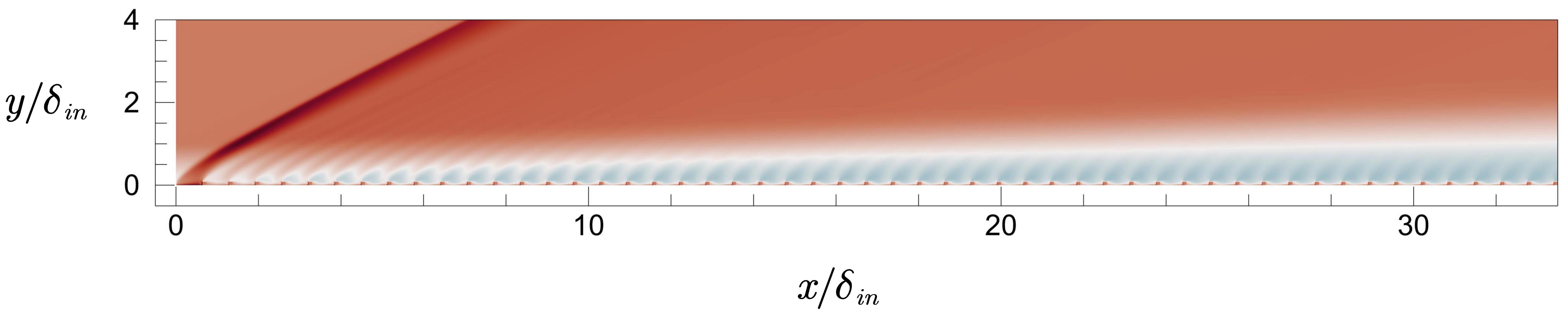}}
            \put(0.01, 0.22){\textbf{(\textit{b})}} 
        \end{picture}
        \label{fig:density_cold}
    \end{subfigure}

    \caption{Contours of Reynolds-averaged density in the rough-wall boundary layers: (a) Adiabatic wall (M2p5RA); (b) Cold wall (M2p5RC).}
    \label{fig:mean_density_contours}
\end{figure}

The transition from the smooth-wall CEBL to the rough-wall CTBL introduces pronounced structural variations in the flow. As shown in Figure \ref{fig:mean_density_contours}, the incoming turbulent boundary layer impinging on the first roughness element triggers a strong leading-edge oblique shock. Downstream of this initial encounter, the subsequent spanwise bars continue to perturb the main flow; however, since the boundary layer has not yet reached a fully equilibrated state with the surface roughness, these bars generate secondary shock waves of relatively lower intensity.

\begin{figure}
    \centering
    \begin{minipage}[t]{0.5\textwidth}
        \raggedright 
        (\textit{a}) \\[0.48em] 
        \includegraphics[width=\textwidth]{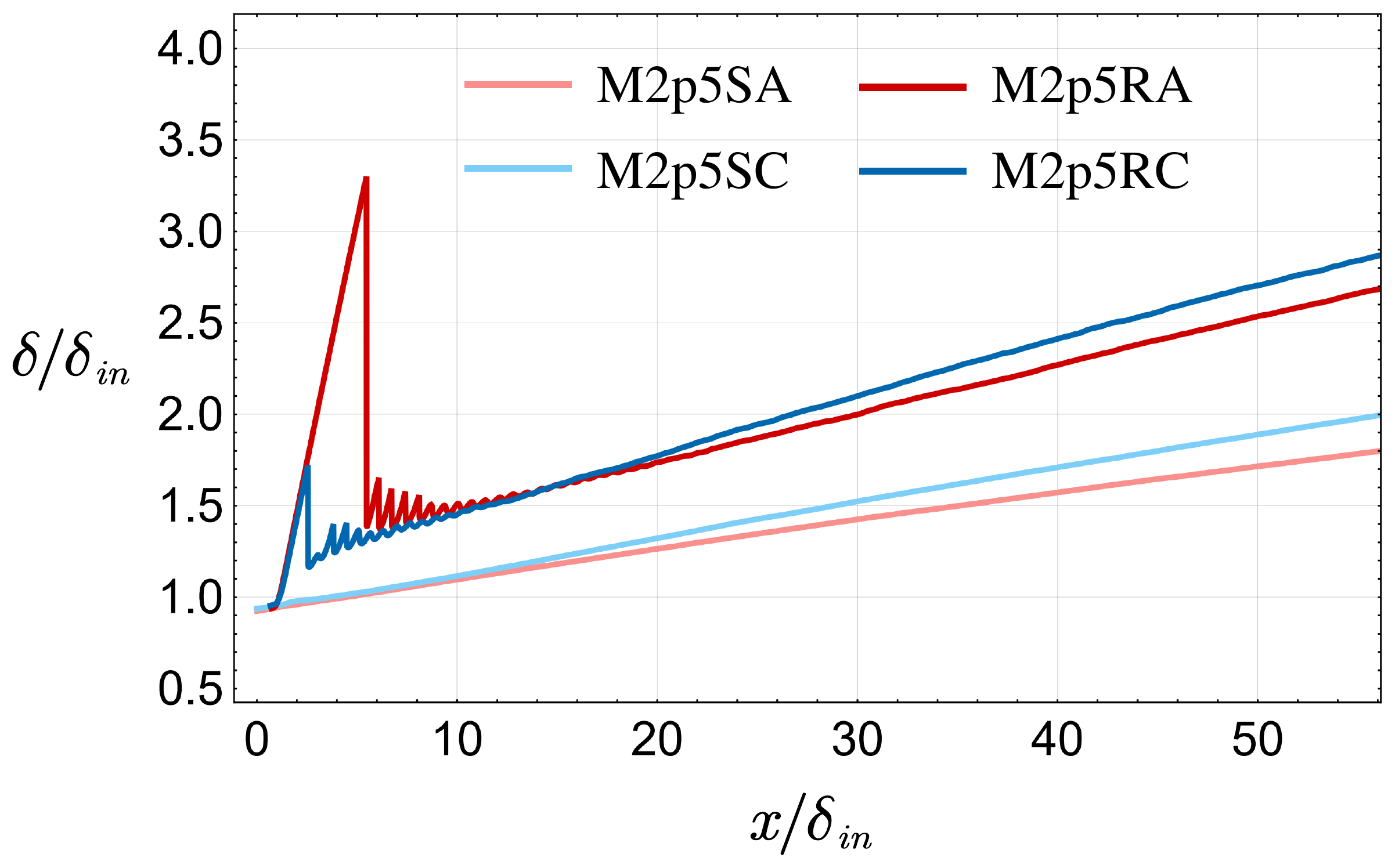}
        \label{fig:x_delta}
    \end{minipage}
    \hfill 
    \begin{minipage}[t]{0.49\textwidth}
        \raggedright 
        (\textit{b}) \\[0.48em] 
        \includegraphics[width=\textwidth]{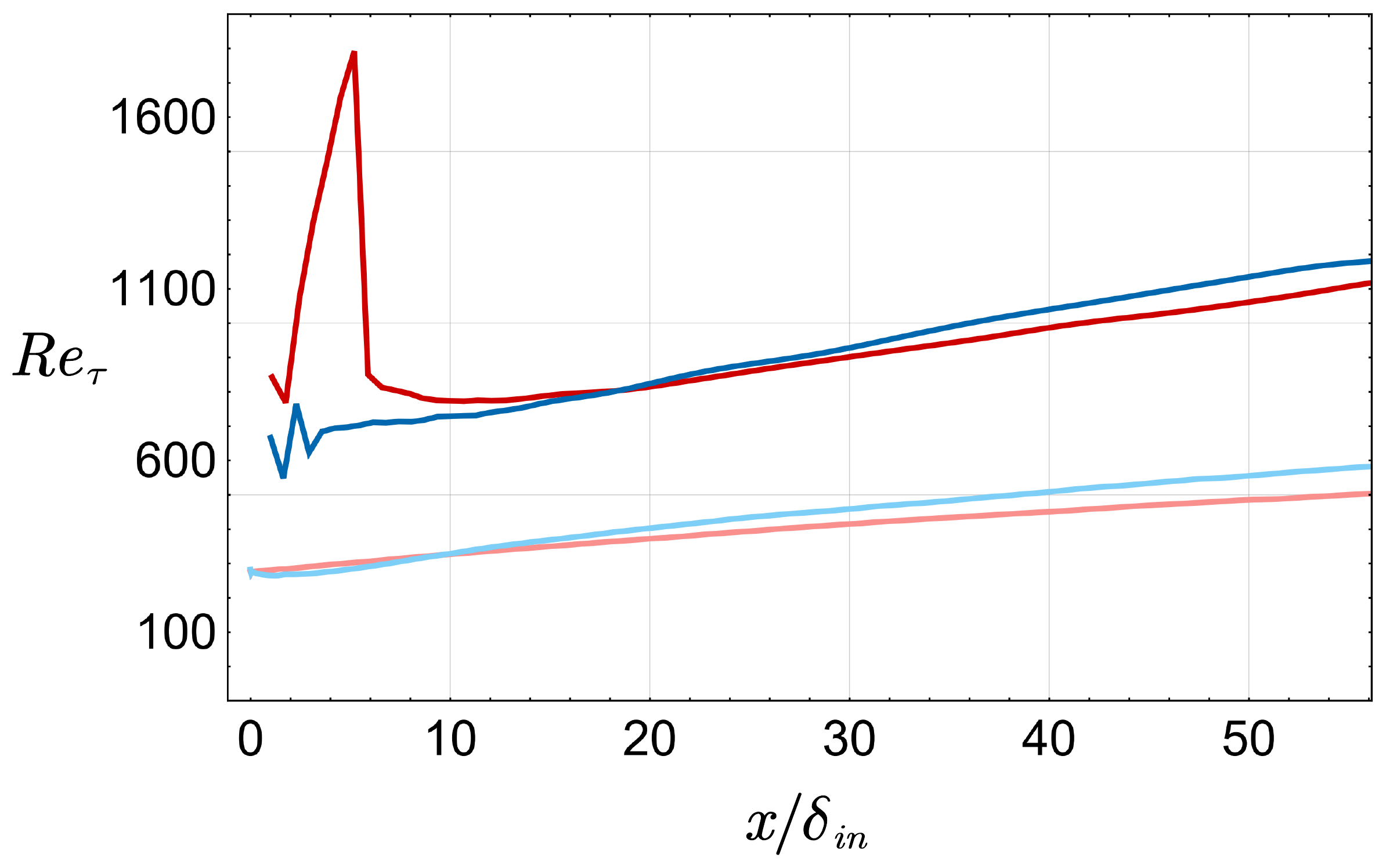} 
        \label{fig:x_ReTau}
    \end{minipage}
    
    \caption{
        Streamwise evolution of (a) boundary layer thickness $\delta$ and (b) friction Reynolds number $Re_\tau$.}
    \label{fig:x_BDL}
\end{figure}

Thermal effects significantly modulate the shock strength. In the adiabatic case (M2p5RA), the higher wall temperature leads to a lower near-wall density, which increases the flow sensitivity to surface irregularities. Consequently, the oblique shocks induced by the downstream bars are notably stronger than those in the cold-wall case (M2p5RC). These roughness-induced shocks gradually weaken as they propagate downstream. By $x = 15\delta_{in}$, the shock-induced perturbations are almost entirely confined within the boundary layer and cease to penetrate into the freestream, a feature also evident in streamwise evolution of the boundary layer thickness shown in Figure \ref{fig:x_BDL}(a). 

Consistent with smooth-wall compressible flows, the cold-wall boundary layer exhibits a more rapid spatial development (Figure \ref{fig:x_BDL}). Owing to the combined effects of the roughness-induced perturbations and the additional form drag, the boundary layer thickness $\delta$ undergoes a rapid expansion immediately following the first roughness element, reaching approximately 1.5 times its smooth-wall counterpart.

\subsection{Drag distribution}\label{sec:Drag_distribution}

\begin{figure}
    \centering
    \includegraphics[width=0.85\textwidth]{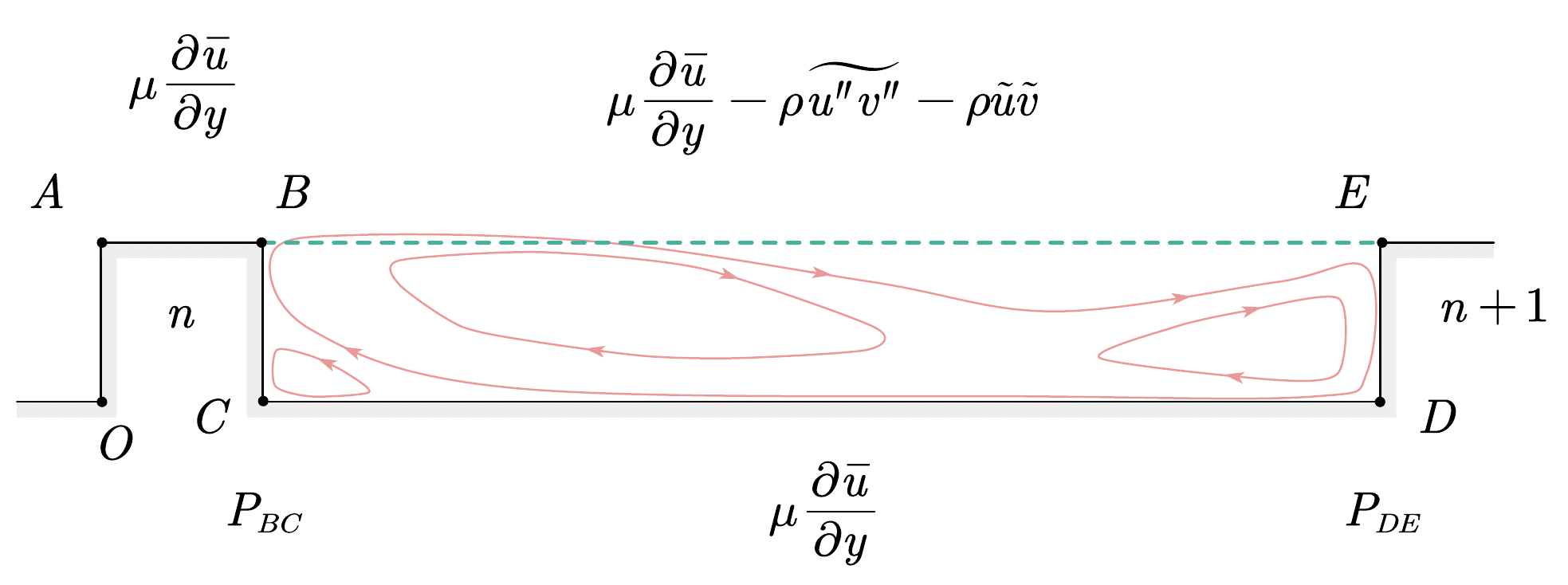}
    
    \caption{
        Schematic of the two-dimensional rough-wall geometry and the decomposition of drag sources. The light red curves represent typical streamlines, traced from the actual flow field around target roughness element in case M2p5RA.
    }
    
    \label{fig:schematic_description}
\end{figure}

The geometric period of the two-dimensional square-rib roughness can be partitioned as illustrated in Figure \ref{fig:schematic_description}. For the $n$-th roughness period, surfaces AB and CD are subjected to frictional forces, contributing to the skin-friction drag. Conversely, the leeward (BC) and windward (ED) faces are subjected to normal stresses, thereby generating the form drag (pressure drag). As detailed in Table \ref{tab:drag_decomposition}, the recirculation zone over the cavity floor (surface CD) yields a negative frictional drag on the order of $-15\%$, whereas the roughness crest (surface AB) contributes a positive frictional drag. Compared to the adiabatic case, the frictional drags on both AB and CD experience a marginal increase under cold-wall conditions, although the variation in their combined relative contribution remains minor. Notably, the form drag constitutes approximately $109\%$ of the total drag.

For the cavity enclosed by BCDE, we consider a virtual plane, BE, situated at the roughness crest. The equivalent drag acting on this virtual plane can be decomposed into three components: the viscous shear stress contribution ($C_{f,BE}$), the Reynolds shear stress contribution ($C_{t,BE}$), and the mean momentum transport ($C_{m,BE}$). According to the principle of momentum conservation, the total drag on plane BE should strictly equal the sum of the drags acting on the physical surfaces BC, CD, and DE. In our evaluation, the residual discrepancy of this balance is bounded within an acceptable margin of approximately $2\%$. At the virtual plane BE, which corresponds to the elevation of the roughness crests, the Reynolds shear stress accounts for a larger fraction of the total stress in the adiabatic case compared to the cold-wall case. By analogy to canonical smooth-wall turbulence, where the Reynolds stress contribution grows with wall-normal distance within the inner layer, the flow state at this virtual plane maps to a higher equivalent smooth-wall $y^+$ for the adiabatic case than for its cold-wall counterpart. Finally, the momentum transport induced by the mean flow ($C_m$) is relatively small. Although it is slightly stronger in the cold-wall case than in the adiabatic case, its magnitude is comparable to the numerical error margin, and thus it is not analyzed in further detail here.

\begin{table}
    \centering
    \begin{tabular}{lcccccc}
        Case & $C_{f,AB}(\%)$ & $C_p(\%)$ & $C_{f,CD}(\%)$ & $C_{f,BE}(\%)$ & $C_{t,BE}(\%)$ & $C_{m,BE}(\%)$ \\[3pt]
        M2p5RA & 5.23 & 109.72 & -14.95 & 11.76 & 79.41 & 1.37 \\
        M2p5RC & 8.66 & 108.50 & -17.16 & 18.81 & 67.21 & 5.00 \\
    \end{tabular}
    \caption{Decomposition of the drag sources for the adiabatic and cold rough-wall cases. The tabulated percentages indicate the relative contributions of the local tangential stresses or momentum transport to the total drag within a single roughness element at $x_t$.}
    \label{tab:drag_decomposition}
\end{table}

\subsection{Mean velocity profiles}\label{sec:Mean_velocity_profiles}

\begin{figure}
    \centering
    \includegraphics[width=0.85\textwidth]{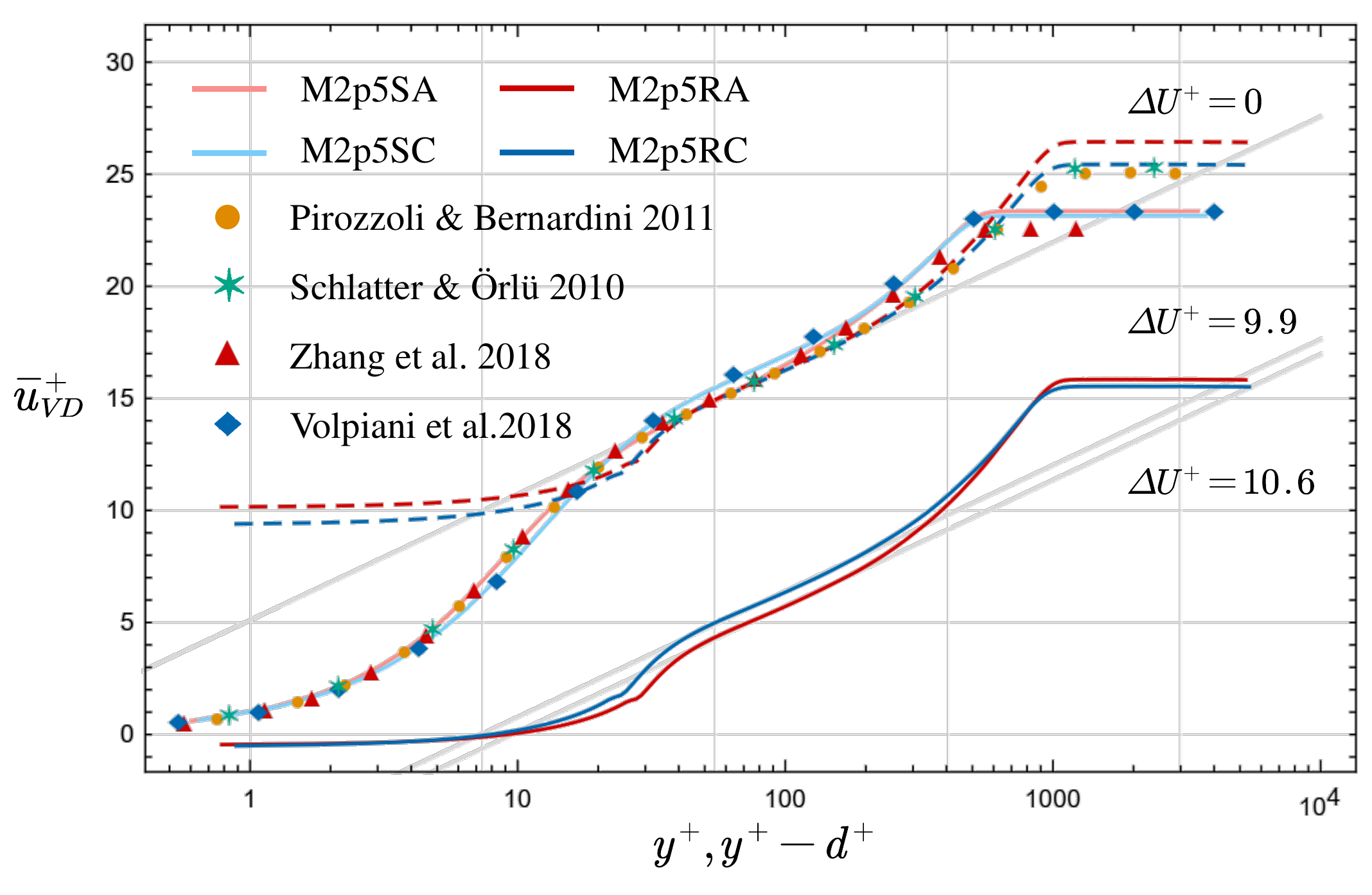}
    \caption{Van Driest (VD) transformation of mean streamwise velocity profiles. Gray lines: logarithmic law. $\Delta U^+$ indicate the downward shift from the smooth-wall case with $\kappa = 0.41$, $B = 5.1$; The dashed lines of corresponding colors: Rough-wall velocity profiles corrected by adding $\Delta U^+$.}
    \label{fig:Mean_VD}
\end{figure}

\begin{figure}
    \centering
    \includegraphics[width=0.85\textwidth]{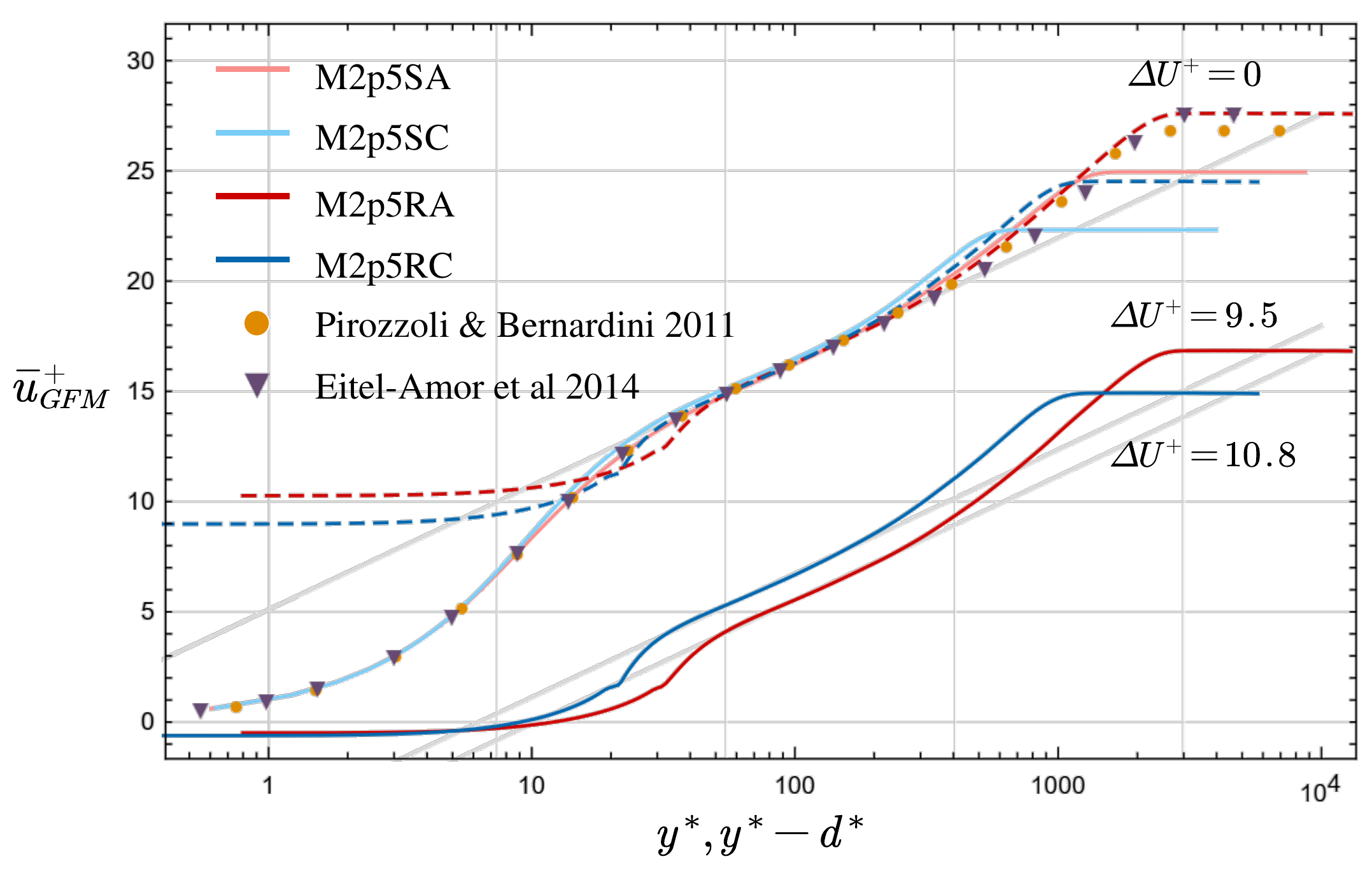}
    \caption{GFM transformation of mean streamwise velocity profiles. Light gray lines: logarithmic law. $\Delta U^+$ indicate the downward shift from the smooth-wall case with $\kappa = 0.41$, $B = 5.1$; The dashed lines of corresponding colors: Rough-wall velocity profiles corrected by adding $\Delta U^+$.}
    \label{fig:Mean_GFM}
\end{figure}

Figure \ref{fig:Mean_VD} reports the mean streamwise velocity profiles for the baseline smooth-wall boundary layers, specifically the adiabatic case (M2p5SA) and the cold-wall case (M2p5SC) at a freestream Mach number of $Ma = 2.5$. As shown in the figure, the present smooth-wall results exhibit excellent agreement with the reference data of \cite{zhang2018direct} ($Ma = 2.5$, $\Rey_{\tau} = 510$, $T_w/T_r=1.0$) and \cite{volpiani2018effects} ($Ma = 2.28$, $\Rey_{\tau} = 511$, $T_w/T_r=0.5$). With the van Driest (VD) transformation applied, the velocity profile of the adiabatic case conforms well to the classical incompressible logarithmic law, whereas the cold-wall case still shows a noticeable deviation due to the significant fluid property variations near the wall. However, when the GFM velocity transformation is applied as shown in Figure \ref{fig:Mean_GFM}, the velocity profiles for both the cold and adiabatic walls collapse satisfactorily onto the incompressible logarithmic line. This not only validates the accuracy of the current numerical setup but also demonstrates the effectiveness of the GFM transformation in accounting for the wall heat flux in smooth boundary layer cases. The incompressible reference cases \cite{schlatter2010assessment} and \cite{eitel2014simulation} are chosen to provide reference profiles for the VD and GFM analyzes, matching the adiabatic rough-wall case (M2p5RA) in terms of the nominal friction Reynolds number ($Re_\tau \approx 1050$) and the semi-local friction Reynolds number ($Re_\tau^* \approx 2480$), respectively. It is noteworthy that the transformed velocity profiles in Figures \ref{fig:Mean_VD} and \ref{fig:Mean_GFM} do not exhibit the discontinuities implied by equation (\ref{eq:spatial_average}). This is because the transformation involves the integration of wall-normal velocity gradients. To avoid numerical singularities arising from the discontinuities in equation (\ref{eq:spatial_average}), a central difference scheme is employed at the roughness crest, yielding a continuous velocity distribution upon integration.

Prior to examining the mean velocity profiles of the rough-wall cases, the zero-plane displacement, $d$, must be rigorously determined, as it is a prerequisite for recovering a well-defined logarithmic region. In the literature, the virtual origin is often defined based on the zero-moment plane of the drag forces acting on the roughness elements (e.g., \cite{jackson1981displacement, lee2007direct}), whereas alternative approaches employ a velocity-profile-fitting methodology (e.g., \cite{cheng2002near, coceal2007structure}). Following the former physical approach, a moment analysis on the two-dimensional roughness elements in the present study yields a zero-moment plane correction of $d_M = 0.60k$ for the adiabatic case (M2p5RA) and $d_M = 0.64k$ for the cold-wall case (M2p5RC). Although these values are highly consistent with the findings of $d_M=0.50$ \citep{lee2011direct} and $d_M=0.62$ \citep{zhang2020rough} in incompressible flow, utilizing $d_M$ as the zero-plane displacement fails to reproduce a satisfactory logarithmic feature. Instead, the profile slope in the expected logarithmic region exhibits an anomalous 'S'-shaped variation (i.e., transitioning from steep to shallow, and back to steep).

\begin{subequations}
    \label{eq:optimization_d}
    \begin{align}
        \mathscr{Y}^{\oplus} &\equiv \ln \left( y^{\oplus}-d^{\oplus} \right), \label{eq:log_coord} \\
        \mathcal{I}(d^{\oplus}) &= \left\{ \mathscr{Y}^{\oplus} : \left| \frac{\mathrm{d} u^{\oplus}}{\mathrm{d} \mathscr{Y}^{\oplus}} - \frac{1}{\kappa} \right| \leqslant \frac{0.1}{\kappa} \right\} = \bigcup_{i\in I} [a_i, b_i], \label{eq:indicator_set} \\
        F(d^{\oplus}) &\equiv \max_{i \in I} \left( b_i - a_i \right), \label{eq:objective_func} \\
        d^{\oplus}_{\mathrm{opt}} &= \underset{d^{\oplus}}{\operatorname{argmax}} F(d^{\oplus}). \label{eq:argmax_d}
    \end{align}
\end{subequations}

To overcome the limitations of the aforementioned moment-based approach and objectively restore the universal logarithmic behavior, a rigorous mathematical optimization procedure is proposed herein. Let $y^{\oplus}$ and $u^{\oplus}$ denote the inner-scaled wall-normal distance and mean streamwise velocity, respectively, under any arbitrary compressibility transformation (e.g., the VD or GFM transformation). The modified logarithmic coordinate is first defined as expressed in (\ref{eq:log_coord}), where $d^{\oplus}$ represents the trial zero-plane displacement in inner scale. Taking the derivative of the velocity with respect to this logarithmic coordinate inherently yields the diagnostic function, $\Xi \equiv \mathrm{d} u^{\oplus} / \mathrm{d} \mathscr{Y}^{\oplus}$. As defined in (\ref{eq:indicator_set}), we identify a set $\mathcal{I}(d^{\oplus})$ comprising all continuous intervals $\bigcup_{i\in I} [a_i, b_i]$, where the index set $I$ labels each distinct interval. This formulation captures all regions wherein the diagnostic function strictly falls within a $\pm 10\%$ tolerance band around the theoretical inverse von Kármán constant, $1/\kappa$. Subsequently, an objective function $F(d^{\oplus})$ is constructed to evaluate the maximum continuous length among these valid logarithmic intervals (\ref{eq:objective_func}). Ultimately, the optimal inner-scaled zero-plane displacement, $d^{\oplus}_{\mathrm{opt}}$, is determined by finding the argument that globally maximizes $F(d^{\oplus})$, as formulated in (\ref{eq:argmax_d}). 

\begin{figure}
    \centering
    \begin{subfigure}{0.987\textwidth}
        \centering
        \setlength{\unitlength}{\textwidth} 
        \begin{picture}(1, 0.25) 
            \put(0.005,0){\includegraphics[width=\textwidth]{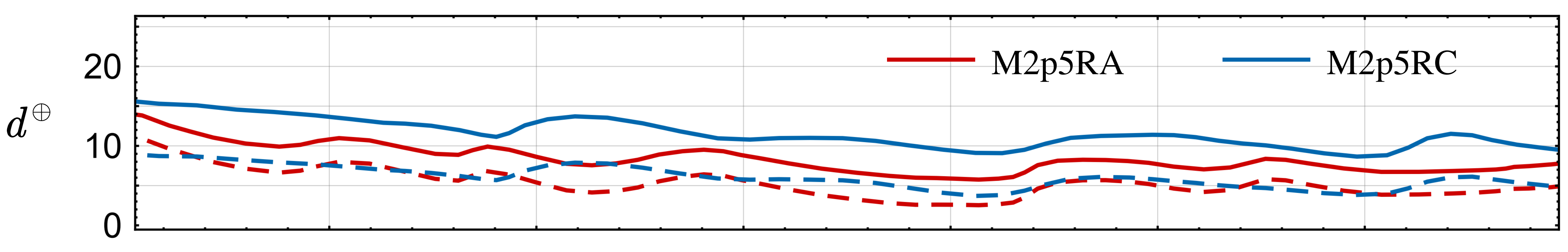}}
            \put(0.01, 0.16){\textbf{(\textit{a})}} 
        \end{picture}
        \label{fig:x_K_ReTau}
    \end{subfigure}

    \vspace{-0.6cm} 

    \begin{subfigure}{\textwidth}
        \centering
        \setlength{\unitlength}{\textwidth}
        \begin{picture}(1, 0.25) 
            \put(0,0){\includegraphics[width=\textwidth]{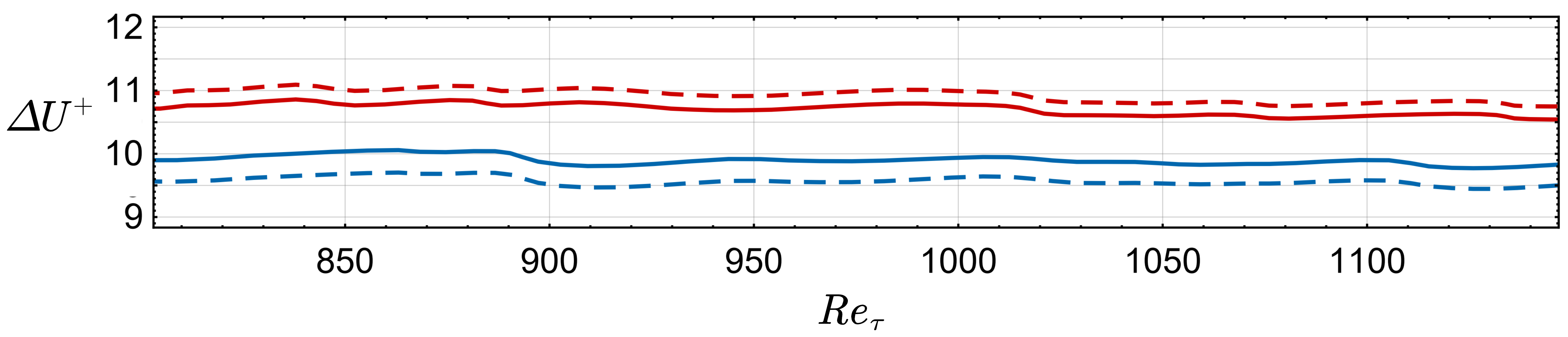}}
            \put(0.01, 0.22){\textbf{(\textit{b})}} 
        \end{picture}
        \label{fig:x_DUP_ReTau}
    \end{subfigure}

    \caption{Evolution of (a) the inner-scaled zero-plane displacement $d^{\oplus}$ and (b) the roughness function $\Delta U^+$ with friction Reynolds number. Solid lines: VD transformation; dashed lines: GFM transformation.}
    \label{fig:displacement_and_shift}
\end{figure}

Table \ref{tab:zero_plane} summarizes the optimal zero-plane displacements determined via the proposed mathematical optimization procedure at target streamwise location $x_t$. Here, $d^+$ and $d^*$ denote the zero-plane displacements normalized by the classical inner scale (associated with the VD transformation) and the GFM inner scale, respectively, whereas $k^+$ and $k^*$ represent the corresponding normalized roughness heights. A comparison between the adiabatic (M2p5RA) and cold-wall (M2p5RC) cases reveals the profound influence of wall heat transfer. In terms of the classical inner scaling, wall cooling noticeably elevates the relative zero-plane displacement, with $d^+/k^+$ increasing from 0.23 to 0.30. However, when the GFM scaling is applied, the absolute values of the effective origin ($d^* \approx 5.7$) collapse remarkably well for both thermal boundary conditions. Consistent behavior is also observed at various streamwise positions for different friction Reynolds numbers (Figure \ref{fig:displacement_and_shift}a). Regardless of the transformation employed, the inner-scaled height between the zero-displacement plane and the roughness crest is invariably greater for the adiabatic wall than for the cold wall. This aligns well with the findings from the drag distribution analysis.

Furthermore, a stark contrast is observed between the kinematically optimized zero-plane displacements ($d^+/k^+$ and $d^*/k^*$, ranging from 0.15 to 0.30) and the estimates derived from the classical zero-moment method ($d_m/k \approx 0.60$--$0.64$). This substantial discrepancy indicates that for the present cavity-type roughness, the aerodynamic centroid of the drag forces is positioned significantly higher than the effective kinematic origin of the overlying boundary layer.

\begin{table}
    \centering
    \begin{tabular}{lccccccccc}
        Case & $d^+$ & $d^*$ & $k^+$ & $k^*$ & $d^+/k^+$ & $d^*/k^*$ & $d_m/k$ & $\Delta U^+$ & $\Delta U^*$\\[3pt]
        M2p5RA & 8.4  & 5.7  & 36.1  & 36.7  & 0.23      & 0.15      & 0.60    & 10.6  & 10.8\\
        M2p5RC & 10.5 & 5.7  & 34.8  & 27.0  & 0.30      & 0.21      & 0.64    & 9.9  & 9.5\\
    \end{tabular}
    \caption{Summary of the normalized roughness heights and zero-plane displacements. $d^+$ and $d^*$ are the optimal zero-plane displacements in classical inner units and GFM inner units at $x_t$, respectively. $d_m/k$ represents the ratio obtained from the classical zero-moment method.}
    \label{tab:zero_plane}
\end{table}

Utilizing the mathematically optimized zero-plane displacements, the mean velocity profiles for the rough-wall cases are evaluated. When scaled by the VD transformation, it is observed that the difference in the freestream velocity, $u^+_{\infty}$, between the adiabatic and cold-wall cases remains marginal. Although the logarithmic intercept of the cold wall is higher than that of the adiabatic wall ($\Delta U^+$ smaller than adiabatic case), the velocity in the wake region increases more rapidly with the wall-normal height for the adiabatic case, a trend consistent with the smooth-wall baselines. 

Conversely, applying the GFM transformation yields a significantly more unified scaling. Under the GFM framework, the difference in the roughness function ($\Delta U^+$) between the cold and adiabatic walls is substantially more pronounced than that observed with the VD transformation. Compared to the incompressible baseline of $\Delta U^+ = 9.86$ \citep{lee2007direct}, the roughness functions for the adiabatic cases, derived from either the VD  or GFM transformation at target streamwise location $x_t$, are consistently higher (Table \ref{tab:zero_plane}). Conversely, for the cold-wall case, the deviations from the incompressible reference are relatively minor, with the GFM transformation yielding a $\Delta U^+ = 9.5$ slightly lower than the incompressible counterpart. In particular, the roughness function $\Delta U^+$ remains nearly independent of $Re_{\tau}$ as it approaches the target Reynolds number range (Figure \ref{fig:displacement_and_shift}b). Furthermore, the shape of the wake-region deviation from the universal log-law is much more consistent between the two thermal conditions. To facilitate a more intuitive assessment of outer-layer similarity, the evaluated velocity deficits $\Delta U^+$ are added back to the respective rough-wall profiles, as denoted by the dashed lines in Figure \ref{fig:Mean_GFM}. Upon correcting for this deficit, the adiabatic rough-wall profile exhibits an excellent collapse with the reference data of \citet{pirozzoli2011turbulence} ($Ma = 2.0$, $\Rey_\tau = 1000$, $T_w/T_r=1.0$) and \citet{eitel2014simulation} (Incompressible $\Rey_\tau = 2478$) in the region strictly above the roughness crest. In the outer wake region, it also maintains a reasonable agreement with \citet{pirozzoli2011turbulence}, while lying slightly above the incompressible reference data of \citet{eitel2014simulation}.

\begin{figure}
    \centering
    
    \begin{minipage}[t]{0.48\textwidth}
        \raggedright 
        (\textit{a}) \\[0.5em] 
        \includegraphics[width=\textwidth]{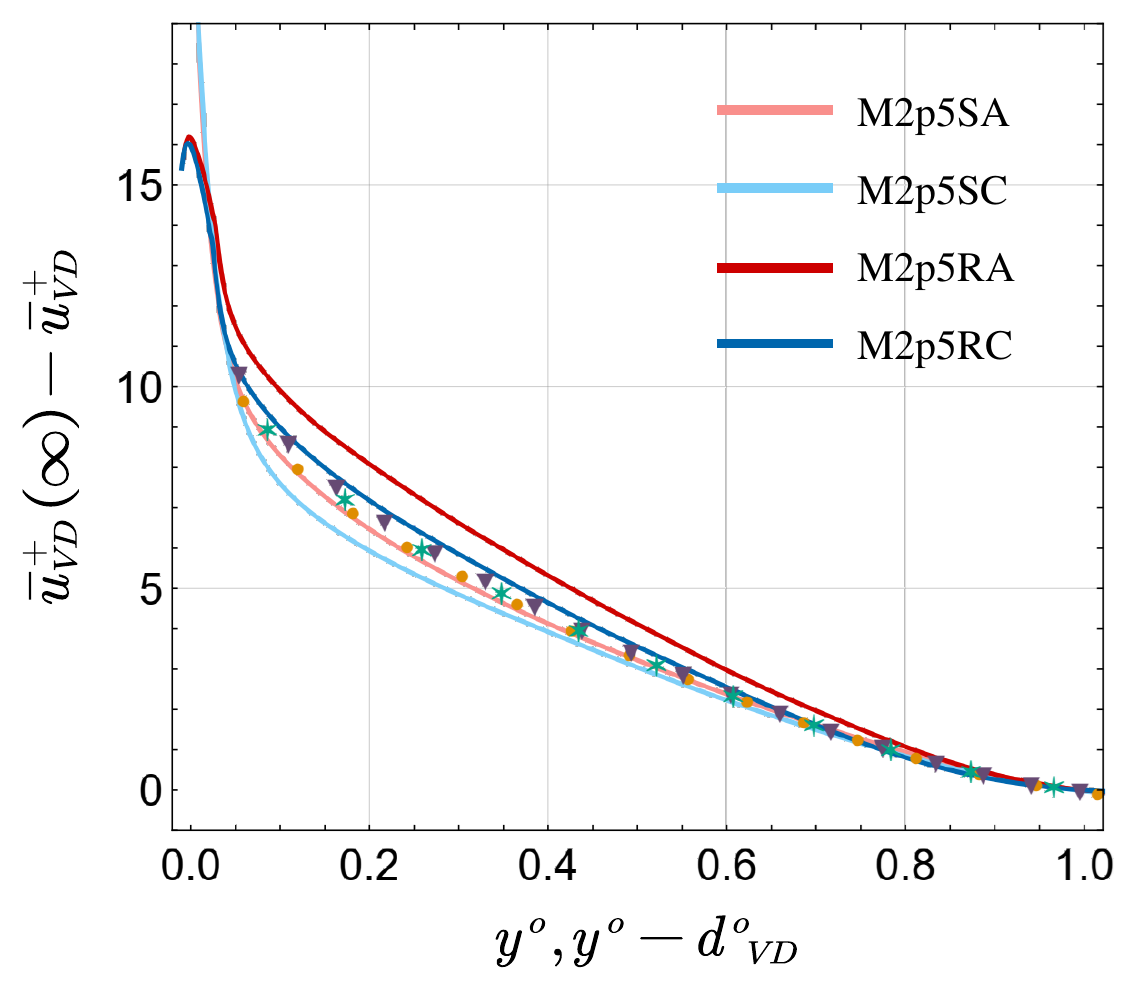}
        \label{fig:Mean_Outer_VD}
    \end{minipage}
    \hfill 
    \begin{minipage}[t]{0.48\textwidth}
        \raggedright 
        (\textit{b}) \\[0.5em] 
        \includegraphics[width=\textwidth]{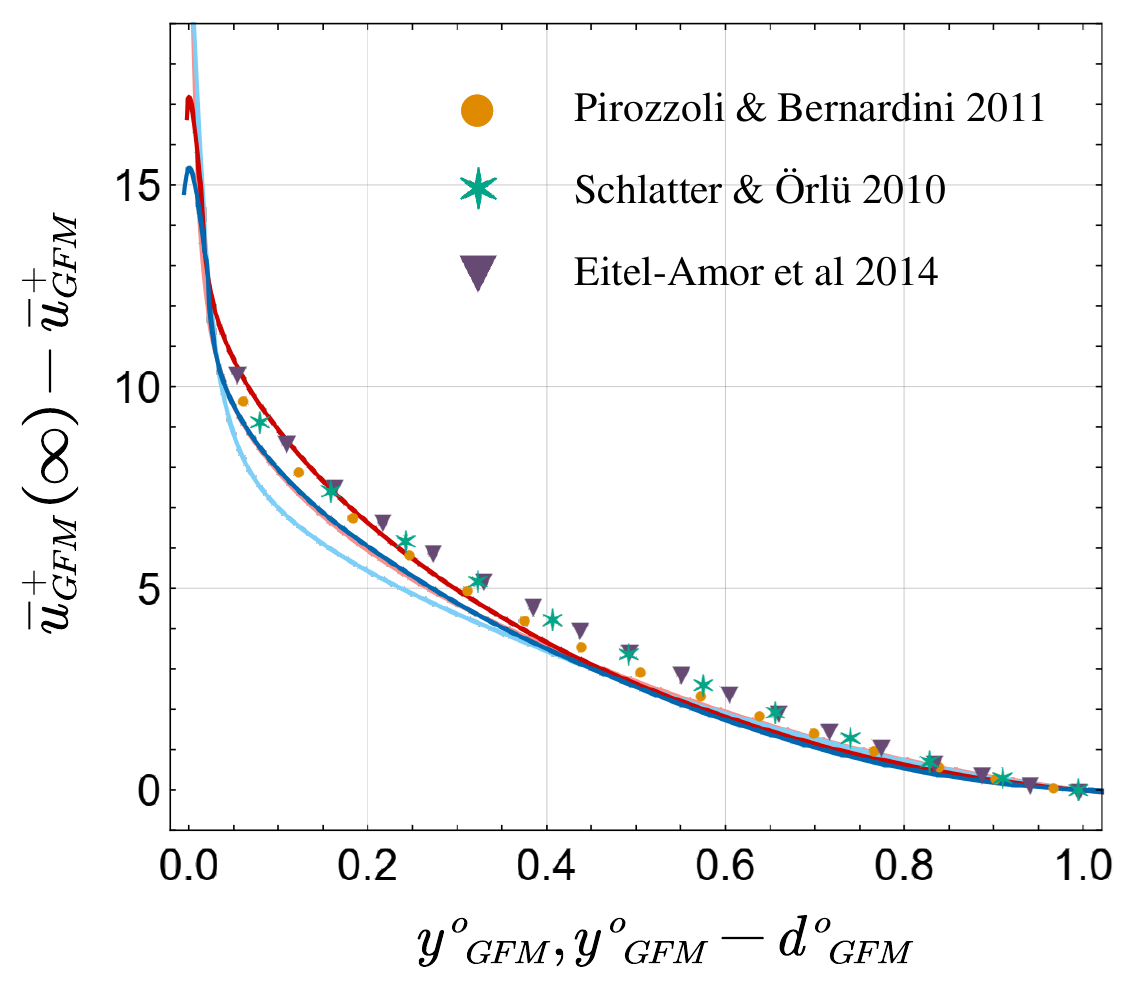} 
        \label{fig:Mean_Outer_GFM}
    \end{minipage}
    
    \caption{Velocity deficit curves: (a) VD transformation; (b) GFM transformation.}
    \label{fig:Mean_Outer_Defect}
\end{figure}

Figure \ref{fig:Mean_Outer_Defect} displays the velocity defect profiles. Here, $y^o$ denotes the corresponding outer-scaled transformed wall-normal coordinate, $d^o$ and $y^o - d^o$ represent the virtual origin displacement and effective wall-normal distance normalized by the effective boundary layer thickness (defined as the distance from the zero-plane displacement to the boundary layer edge in transformed wall-normal coordinate). As observed in Figure \ref{fig:Mean_Outer_Defect}(a), under the VD transformation, substantial discrepancies exist among the outer-region velocity profiles, indicating an absence of outer-layer similarity. Conversely, when the GFM transformation is applied (Figure \ref{fig:Mean_Outer_Defect}(b)), the four $Ma = 2.5$ cases, encompassing both smooth and rough walls, as well as adiabatic and cold-wall conditions---exhibit a good collapse in the region of $y^o > 0.4$. This collapse demonstrates the outer-layer similarity of the velocity defect under the GFM transformation, which is consistent with the incompressible findings of \citet{lee2007direct}. 

However, discernible differences remain when comparing the present M2p5 series with the reference data of \citet{pirozzoli2011turbulence} and incompressible cases \citep{schlatter2010assessment, eitel2014simulation} in Figure \ref{fig:Mean_Outer_Defect}(b). Specifically, the $Ma = 2.0$ case of \citet{pirozzoli2011turbulence} shows a relatively minor deviation, with a noticeable departure emerging below $y^o < 0.6$. In contrast, the discrepancy is more pronounced when compared against the incompressible results of \citet{schlatter2010assessment} and \citet{eitel2014simulation}. These observations suggest that the shape of the wake region under the GFM transformation inherently varies with the freestream Mach number. Nevertheless, for a given Mach number, the transformation successfully maps the wake profile of a rough wall to a shape consistent with the smooth-wall baseline, even in the presence of wall heat transfer and disparate friction Reynolds numbers ($\Rey_{\tau}$).

\subsection{Velocity fluctuations}\label{sec:Velocity_fluctuations}

\begin{figure}
    \centering
    \includegraphics[width=0.85\textwidth]{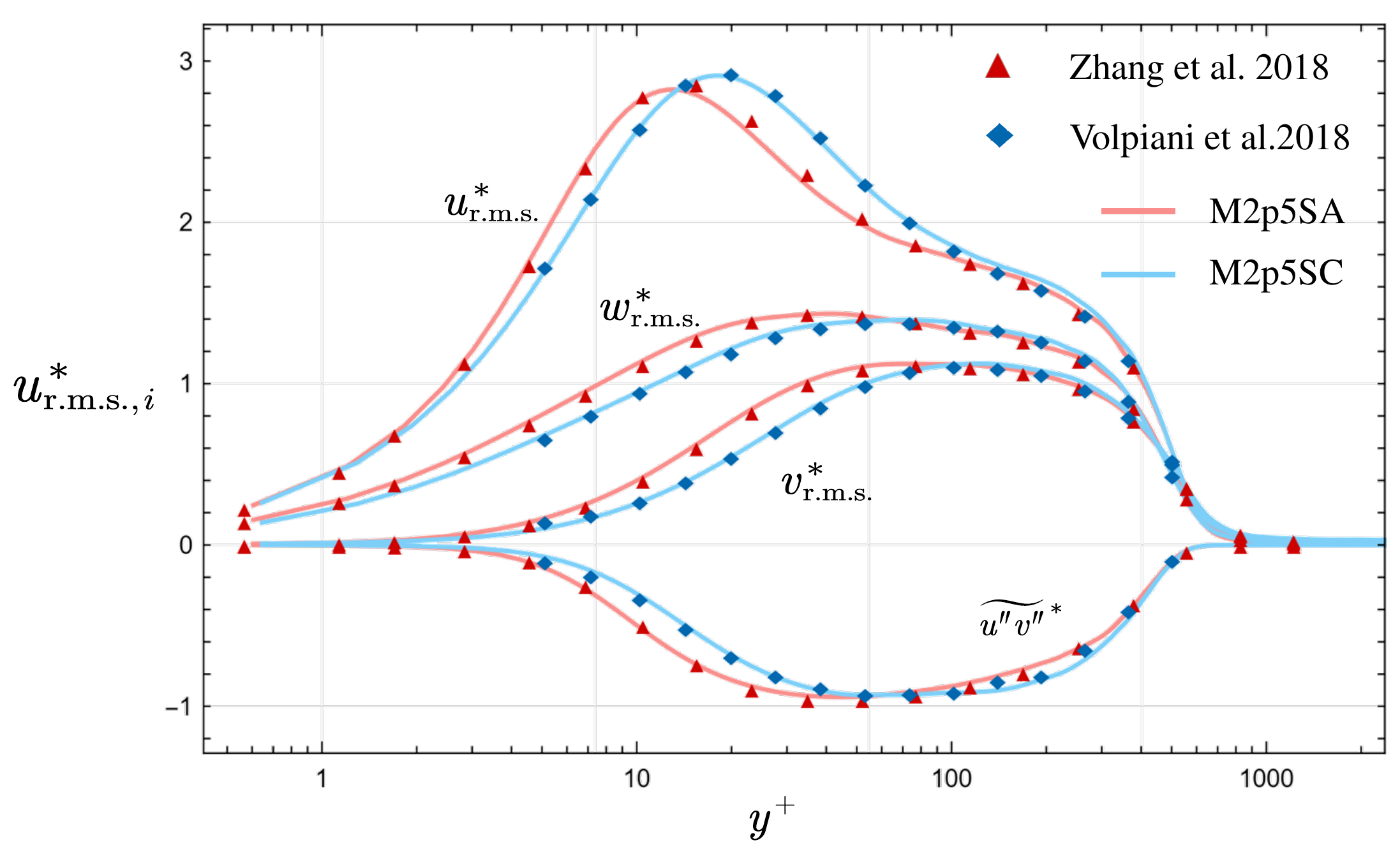}
    \caption{Turbulent velocity fluctuations of smooth-wall cases scaled according to equation (\ref{eq:rms_scaling}).}
    \label{fig:RMS_Smooth}
\end{figure}

Figure \ref{fig:RMS_Smooth} displays the wall-normal distributions of the root-mean-square (r.m.s.) velocity fluctuations for the baseline smooth boundary layers. To account for the mean density variations induced by compressibility, the velocity fluctuations are scaled according to \cite{morkovin1962effects}, defined as 
\begin{equation}
    u_{r.m.s.,i}^{*}=\left( \frac{\bar{\rho}}{\bar{\rho}_w} \right) ^{1/2}\frac{\sqrt{\widetilde{u_{i}^{\prime \prime}u_{i}^{\prime \prime}}}}{u_{\tau}}
    \label{eq:rms_scaling}
\end{equation}
As depicted in the figure, the density-scaled r.m.s. velocity profiles for both the adiabatic and cold-wall smooth cases exhibit good agreement with the reference DNS results of \citet{zhang2018direct} and \citet{volpiani2018effects}. This validation further confirms the reliability of the present numerical setup.

\begin{figure}
    \centering
    
    \begin{minipage}[t]{0.48\textwidth}
        \raggedright 
        (\textit{a}) \\[0.5em] 
        \includegraphics[width=\textwidth]{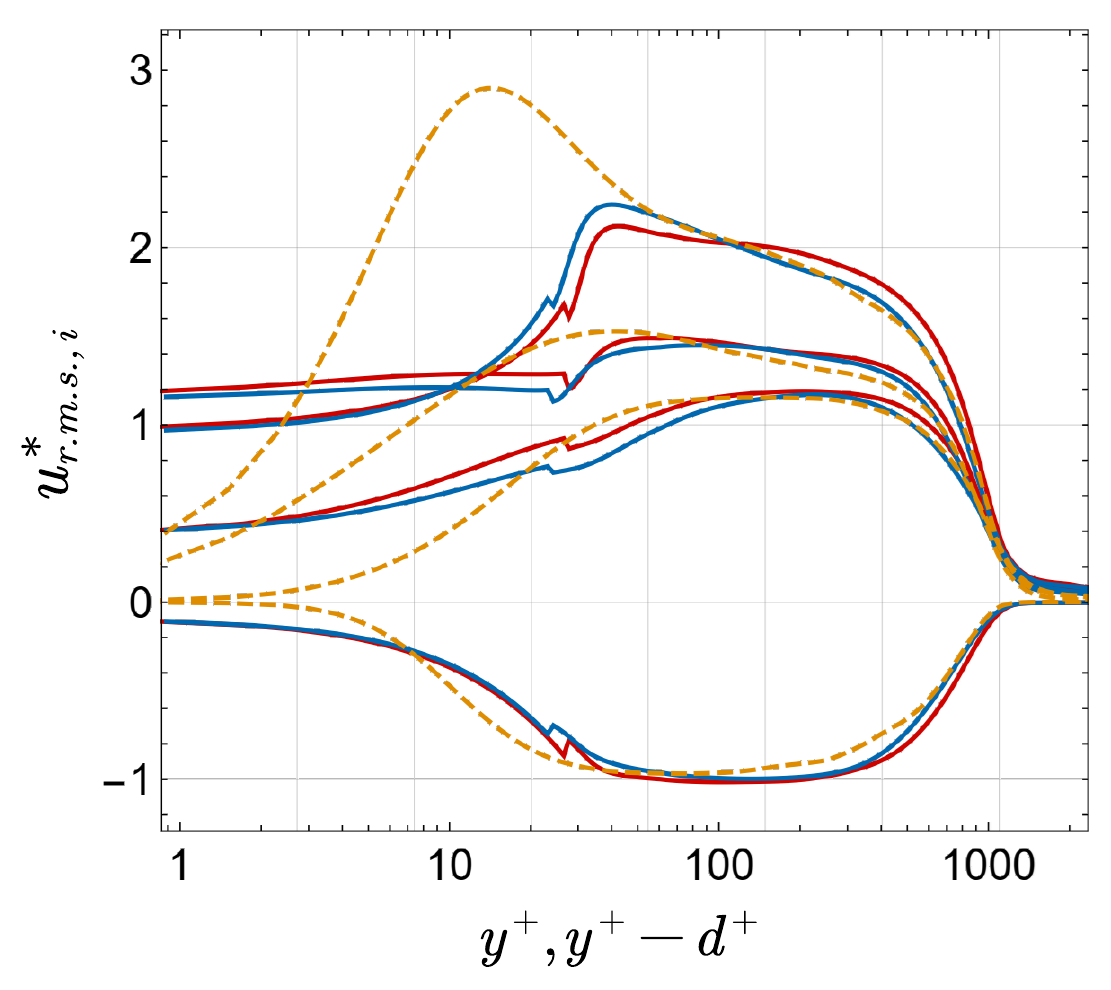}
        \label{fig:RMS_Rough_Inner}
    \end{minipage}
    \hfill 
    \begin{minipage}[t]{0.48\textwidth}
        \raggedright 
        (\textit{b}) \\[0.5em] 
        \includegraphics[width=\textwidth]{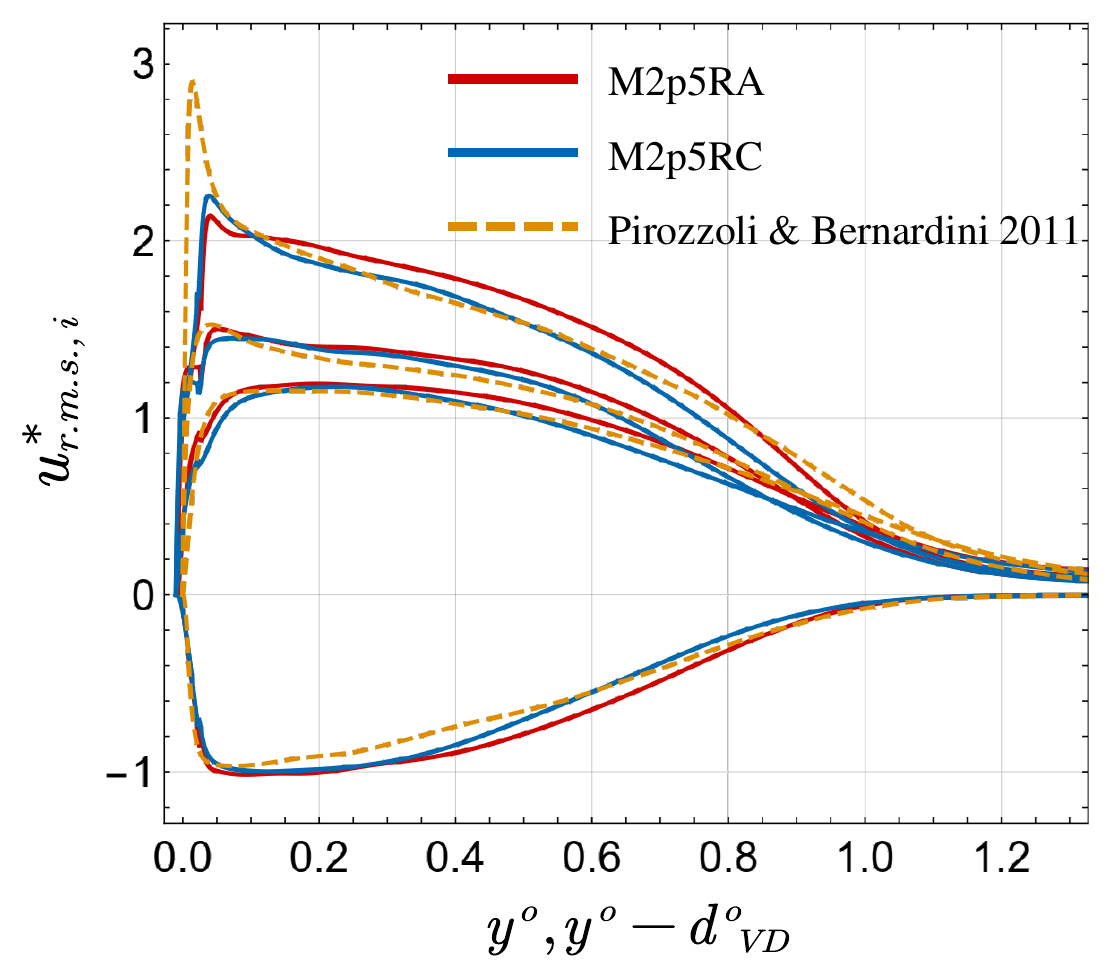} 
        \label{fig:RMS_Rough_Outer}
    \end{minipage}
    
    \caption{Turbulent velocity fluctuations scaled according to equation (\ref{eq:rms_scaling}) as a function of $y$ normalized by (a) inner scale, and (b) outer scale.}
    \label{fig:RMS_Rough}
\end{figure}

For the rough-wall boundary layers, the peak of the streamwise velocity fluctuation is displaced away from the wall to a location above the roughness crest, and its magnitude is significantly attenuated compared to the smooth-wall baseline (Figure \ref{fig:RMS_Rough}a). Concurrently, the overall fluctuation intensities in the inner region are lower than those of the smooth-wall counterparts. Comparing the adiabatic (M2p5RA) and cold-wall (M2p5RC) cases in Figure \ref{fig:RMS_Rough}(a), the cold-wall streamwise fluctuation, $u^*_{r.m.s.}$, is stronger in the near-wall region, whereas the wall-normal and spanwise components ($v^*_{r.m.s.}$ and $w^*_{r.m.s.}$) are weaker. This thermal dependence is consistent with the trend observed in the smooth-wall cases in Figure \ref{fig:RMS_Smooth}.

In the outer region, however, the behavior of the rough-wall fluctuations is more complex. In the lower portion of the outer region ($0.1 < y^o < 0.6$), the fluctuation intensities of the rough-wall turbulence consistently exceed those of the smooth boundary layers, exhibiting a slower rate of decay in the wall-normal direction. Conversely, towards the edge of the boundary layer ($0.6 < y^o < 1$), the rough-wall fluctuations decay abruptly. Notably, the onset of this rapid attenuation occurs at a lower wall-normal height for the cold-wall case compared to the adiabatic case. Throughout the entire outer region, the fluctuation intensities of the adiabatic wall remain stronger than those of the cold wall. Such a rapid decay near the boundary layer edge has also been reported in incompressible rough-wall boundary layers by \citet{lee2011direct}. Overall, the pronounced discrepancy between the rough- and smooth-wall velocity fluctuation profiles in the outer region suggests an absence of strict outer-layer similarity for the turbulence intensities.

\subsection{General Reynolds analogy}\label{sec:GRA}

\begin{figure}
    \centering
    \begin{minipage}[t]{0.48\textwidth}
        \raggedright 
        (\textit{a}) \\[0.5em] 
        \includegraphics[width=\textwidth]{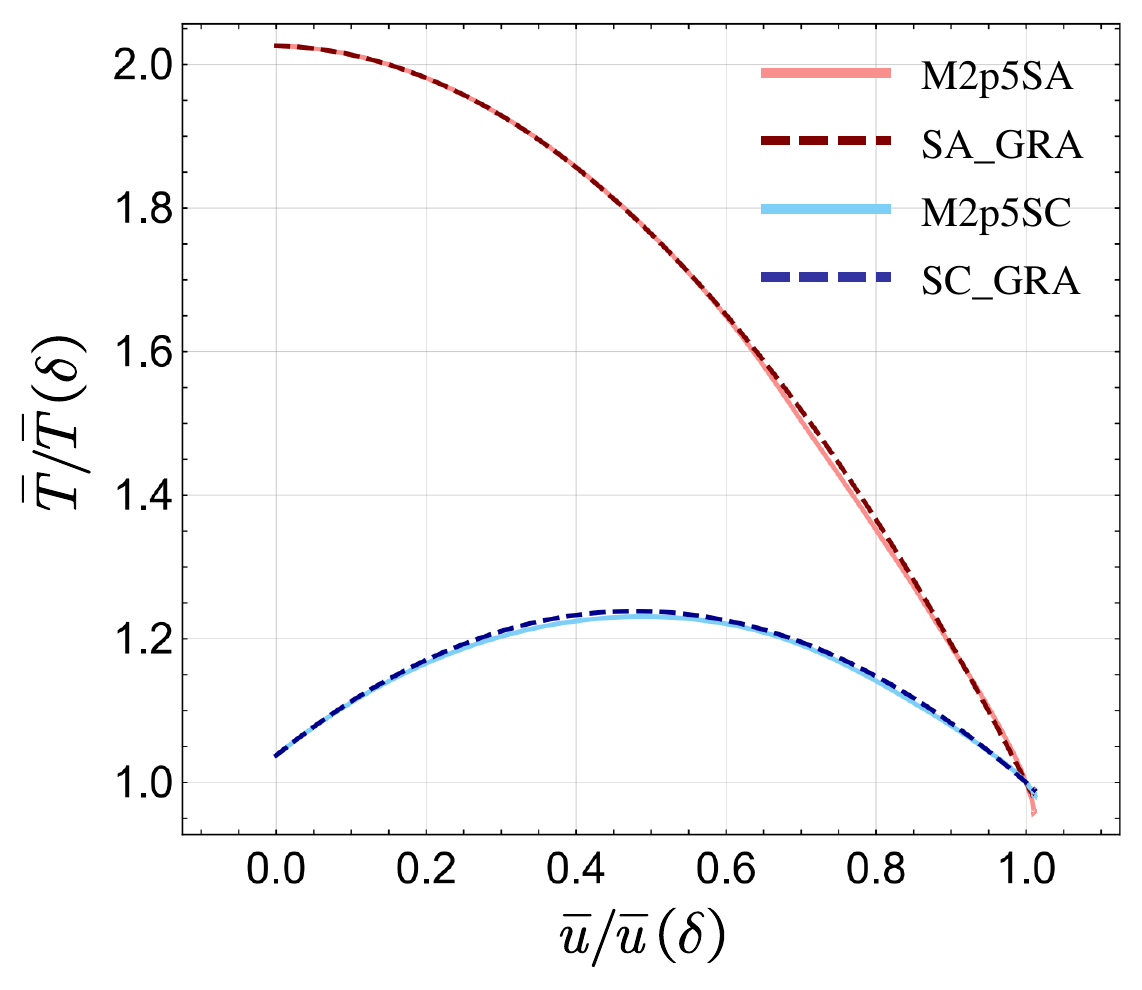}
        \label{fig:GRA_smooth}
    \end{minipage}
    \hfill 
    \begin{minipage}[t]{0.48\textwidth}
        \raggedright 
        (\textit{b}) \\[0.5em] 
        \includegraphics[width=\textwidth]{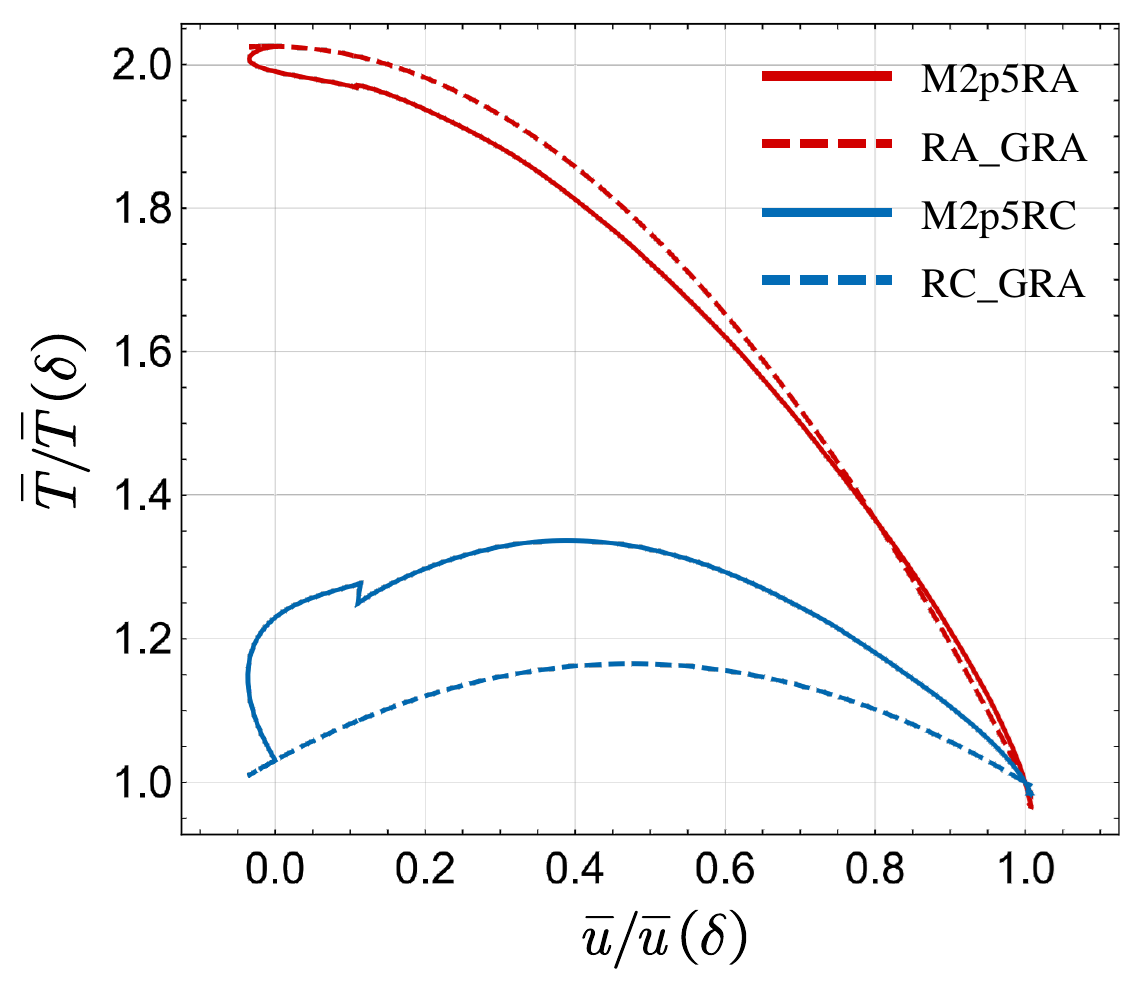} 
        \label{fig:GRA_inner}
    \end{minipage}
    
    \caption{Mean temperature-velocity relations in outer scaling: (a) Smooth-wall cases; (b) rough-wall cases. Solid lines: DNS. Dashed lines: GRA prediction.}
    \label{fig:GRA_Classic}
\end{figure}

For the mean temperature field, the theoretical distributions can be reconstructed using the generalized Reynolds analogy (GRA) formulated by \citet{zhang2014generalized}. As shown in Figure \ref{fig:GRA_Classic}(a) for the smooth-wall baselines, the temperature-velocity relations predicted by the GRA show good agreement with the present DNS results. Conversely, discrepancies arise in the rough-wall cases (Figure \ref{fig:GRA_Classic}(b)). For the adiabatic rough wall (M2p5RA), the GRA-reconstructed temperature profile deviates from the actual simulation data primarily in the near-wall inner region, with the error gradually diminishing as the wall distance increases. This observation aligns with the findings of \citet{cogo2025development}.

The velocity-temperature relationship becomes considerably more complex for the cold-wall rough case (M2p5RC), particularly below the roughness crest. The actual simulation reveals a sharp increase in temperature accompanied by only marginal variations in velocity. However, the ratio of the actual wall heat flux to the skin friction, which dictates the slope of the GRA curve at the origin, is insufficient to capture this abrupt change. As a result, the temperature curve predicted by the GRA lies significantly below the actual distribution. Although the discrepancy narrows in the outer region, this convergence is primarily because the GRA calculation utilizes the freestream conditions at the boundary layer edge as the reference point. Consequently, the classical GRA is inadequate for describing rough-wall boundary layers subjected to substantial wall heat transfer.

To accurately reconstruct the temperature profiles over rough walls, a more rigorous scrutiny of the applicability of GRA is required. \citet{zhang2014generalized} derived that the recovery factor, $r_g$, at an arbitrary height in a smooth-wall turbulent boundary layer satisfies 
\begin{equation}
    r_g=\frac{C_p}{\bar{u}}\left[ \Gamma_w-\frac{1}{Pr_e}\frac{\partial \bar{T}}{\partial \bar{u}} \right]
    \label{eq:recovery_smooth}
\end{equation}
The dimensional parameter $\Gamma _w\equiv \left( -Pr/C_p \right) \left( \overline{q}_w/\overline{\tau }_w \right) $ based on surface fluxes characterizes the wall thermo-kinematic relationship, naturally recovering the gradient $\left( \partial \overline{T}/\partial \overline{u} \right) _w$ for smooth walls. Consequently, the velocity-temperature relation within the boundary layer is governed by the differential equation
\begin{equation}
    \bar{T}-\frac{\bar{u}}{2}\left[ \Gamma_w+\frac{1}{Pr_e}\frac{\partial \bar{T}}{\partial \bar{u}} \right] =\bar{T}_w
    \label{eq:diff_vel_temp}
\end{equation}
Empirical evidence suggests that the effective turbulent Prandtl number, $Pr_e$, is approximately unity throughout the boundary layer. Integrating this differential equation and applying three boundary conditions, specifically the Dirichlet and Neumann conditions at the wall and the Dirichlet condition at the freestream edge, yields the classical GRA relation.
\begin{equation}
    \frac{\bar{T}}{\bar{T}_\delta}=\frac{\bar{T}_w}{\bar{T}_\delta}+\left(\frac{\bar{u}_\delta}{\bar{T}_\delta} \Gamma_w\right)\left(\frac{\bar{u}}{\bar{u}_\delta}\right)+\left(\frac{\bar{T}_\delta-\bar{T}_w}{\bar{T}_\delta}-\frac{\bar{u}_\delta}{\bar{T}_\delta} \Gamma_w\right)\left(\frac{\bar{u}}{\bar{u}_\delta}\right)^2
    \label{eq:classic_GRA}
\end{equation}
For rough walls, constructing a model based on outer-layer similarity necessitates a simplification of the physical properties within the roughness sublayer. For instance, \citet{fu2023effects} employed a Robin boundary condition to model densely packed square ribs in numerical simulations. In the present study, to recover the GRA in the outer region, we postulate the existence of an equivalent slip plane located below the roughness crest. By assigning a slip velocity $u_w$ and virtual boundary conditions to this plane, the outer-region temperature profile can be reconstructed. By analogy, the modified recovery factor satisfies
\begin{equation}
    r_g=\frac{C_p}{\bar{u}-u_w}\left[ \Gamma_w-\frac{1}{Pr_e}\frac{\partial \bar{T}}{\partial \bar{u}} \right]
    \label{eq:recovery_rough}
\end{equation}
which in turn leads to the differential velocity-temperature relation expressed as
\begin{equation}
    \bar{T}-\frac{\bar{u}-u_w}{2}\left[ \Gamma_w+\frac{1}{Pr_e}\frac{\partial \bar{T}}{\partial \bar{u}} \right] =\bar{T}_w
    \label{eq:diff_vel_temp_rough}
\end{equation}

From a practical standpoint, directly integrating from this virtual wall to obtain the temperature profile and subsequently fitting it to the actual outer-region velocity-temperature relation significantly increases both the mathematical complexity and application difficulty. An alternative approach is to introduce the information of a reference point within the outer region to serve directly as a physical constraint, thereby constructing the quadratic velocity-temperature relation. We demonstrate herein that these two methodologies are mathematically equivalent. Consider an arbitrary reference point $i$ selected for the reference-point method, and another arbitrary outer-region point $j$ chosen to establish equivalence. Under the framework of the virtual-wall method, the differential velocity-temperature relation (\ref{eq:diff_vel_temp_rough}) evaluated at these points yields
\begin{subequations}
\label{eq:F_ab}
\begin{align}
    \bar{T}_i-\bar{T}_w=\frac{\bar{u}_i-u_w}{2}\left[ \Gamma_w+\left. \frac{\partial \bar{T}}{\partial \bar{u}} \right|_i \right] \label{eq:Fa} \\
    \bar{T}_j-\bar{T}_w=\frac{\bar{u}_j-u_w}{2}\left[ \Gamma_w+\left. \frac{\partial \bar{T}}{\partial \bar{u}} \right|_j \right] \label{eq:Fb}
\end{align}
\end{subequations}
Subtracting (\ref{eq:Fa}) from (\ref{eq:Fb}) and exploiting the quadratic nature of the differential equation's solution leads directly to 
\begin{equation}
    \begin{aligned}
        \bar{T}_j-\bar{T}_i &= \frac{\bar{u}_j-u_w}{2}\left[ \Gamma_w+\left. \frac{\partial \bar{T}}{\partial \bar{u}} \right|_j \right] -\frac{\bar{u}_i-u_w}{2}\left[ \Gamma_w+\left. \frac{\partial \bar{T}}{\partial \bar{u}} \right|_i \right] \\
        &= \frac{\bar{u}_j-\bar{u}_i}{2}\left[ \left. \frac{\partial \bar{T}}{\partial \bar{u}} \right|_i+\left. \frac{\partial \bar{T}}{\partial \bar{u}} \right|_j \right]
    \end{aligned}
    \label{eq:method2_base}
\end{equation}
which constitutes the fundamental equation of the reference-point method. Thus, the virtual-wall integration and the reference-point method are strictly equivalent. Consequently, for any virtual or reference height $h$, the generalized Reynolds analogy can be recast as equation (\ref{eq:GRA_reference_h}). Hereafter, this rough-wall formulation of the generalized Reynolds analogy is referred to as the rGRA.
\begin{equation}
    \begin{aligned}
        \bar{T}_{\mathrm{rGRA}}\left( h,\bar{u} \right)=\bar{T}_h+\left( \left. \frac{\partial \bar{T}}{\partial \bar{u}} \right|_h \right) \left( \bar{u}-u_h \right) +\left( \bar{T}_{\delta}-\bar{T}    _h-\left. \left( \bar{u}_{\delta}-u_h \right) \frac{\partial \bar{T}}{\partial \bar{u}} \right|_h \right) \left( \frac{\bar{u}-u_h}{\bar{u}_{\delta}    -u_h} \right) ^2
    \end{aligned}
    \label{eq:GRA_reference_h}
\end{equation}
It is worth noting that this virtual-wall formulation is inherently under-determined. Once a slip velocity is prescribed, the corresponding virtual temperature and the ratio of heat flux to skin friction can be adjusted to best fit the outer-region velocity-temperature distribution. If the slip velocity is set to zero, the resulting virtual boundary conditions hold no \textit{a priori} theoretical connection to the true physical conditions at the rough wall; rather, they represent the optimal mathematical solution that satisfies the velocity-temperature relation.

\begin{figure}
    \centering
    \begin{minipage}[t]{0.48\textwidth}
        \raggedright 
        (\textit{a}) \\[0.5em] 
        \includegraphics[width=\textwidth]{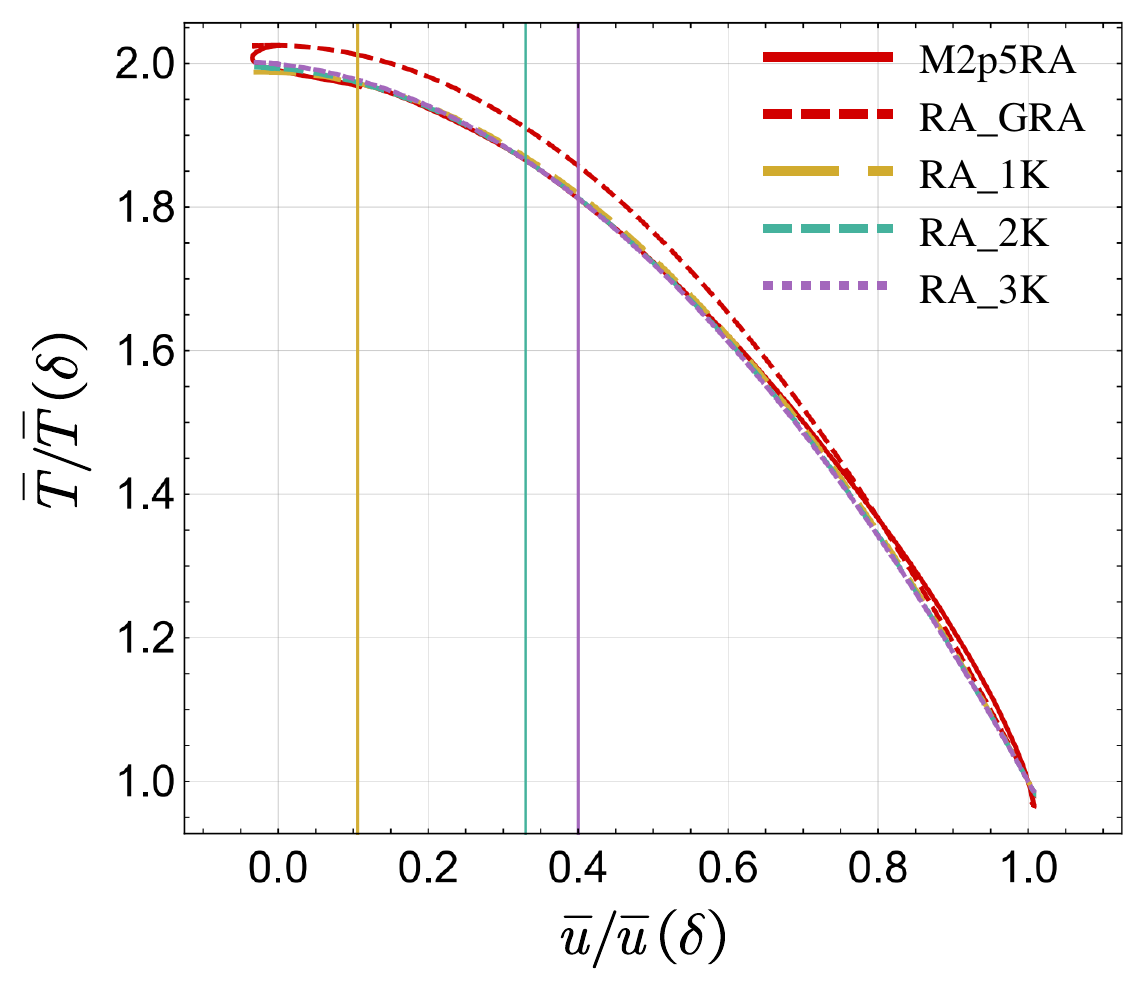}
        \label{fig:GRA_New_A}
    \end{minipage}
    \hfill 
    \begin{minipage}[t]{0.48\textwidth}
        \raggedright 
        (\textit{b}) \\[0.5em] 
        \includegraphics[width=\textwidth]{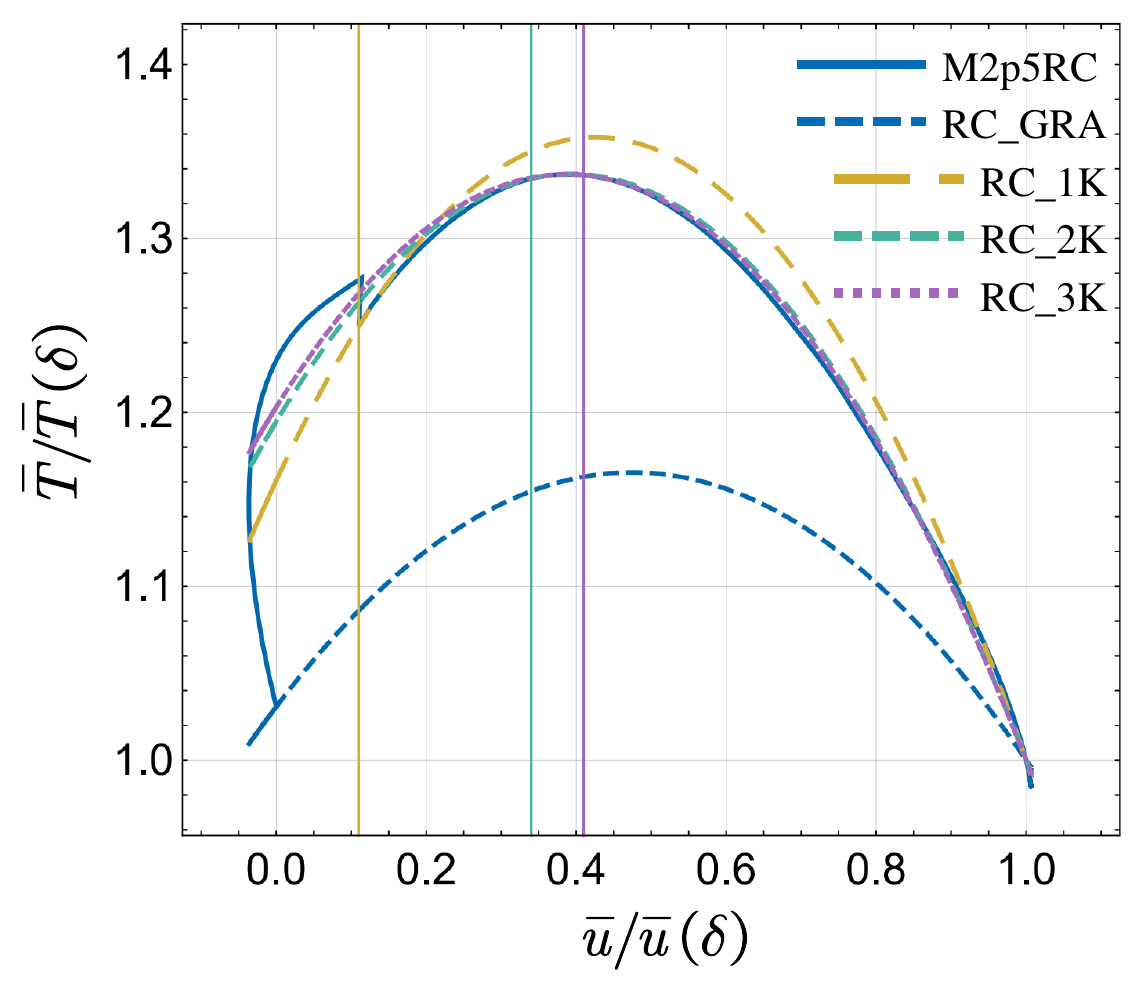} 
        \label{fig:GRA_New_C}
    \end{minipage}
    
    \caption{Mean temperature-velocity relations in outer scaling: (a) Adiabatic rough-wall case; (b) cold rough-wall case. Solid lines: DNS. Dashed lines of the same colors with solid lines: GRA prediction. Yellow dashed lines: $\bar{T}_{\mathrm{rGRA}}(k,\bar{u})$; Green dashed lines: $\bar{T}_{\mathrm{rGRA}}(2k,\bar{u})$; and Purple dashed lines:   $\bar{T}_{\mathrm{rGRA}}(3k,\bar{u})$. Vertical solid lines: Mean velocities corresponding to the reference heights $ik$ selected for the rGRA.}
    \label{fig:GRA_New}
\end{figure}

To reconstruct the temperature-velocity relations using the rGRA, we select three reference heights: $k$, $2k$, and $3k$, spanning the region from the roughness crest to the upper edge of the roughness sublayer. For the adiabatic rough-wall case (Figure \ref{fig:GRA_New}a), the reconstructed profiles anchored at all three reference heights exhibit good agreement with the DNS results. This agreement is observed almost immediately above the roughness crest. 

\begin{figure}
    \centering
    \begin{minipage}[t]{0.48\textwidth}
        \raggedright 
        (\textit{a}) \\[0.5em] 
        \includegraphics[width=\textwidth]{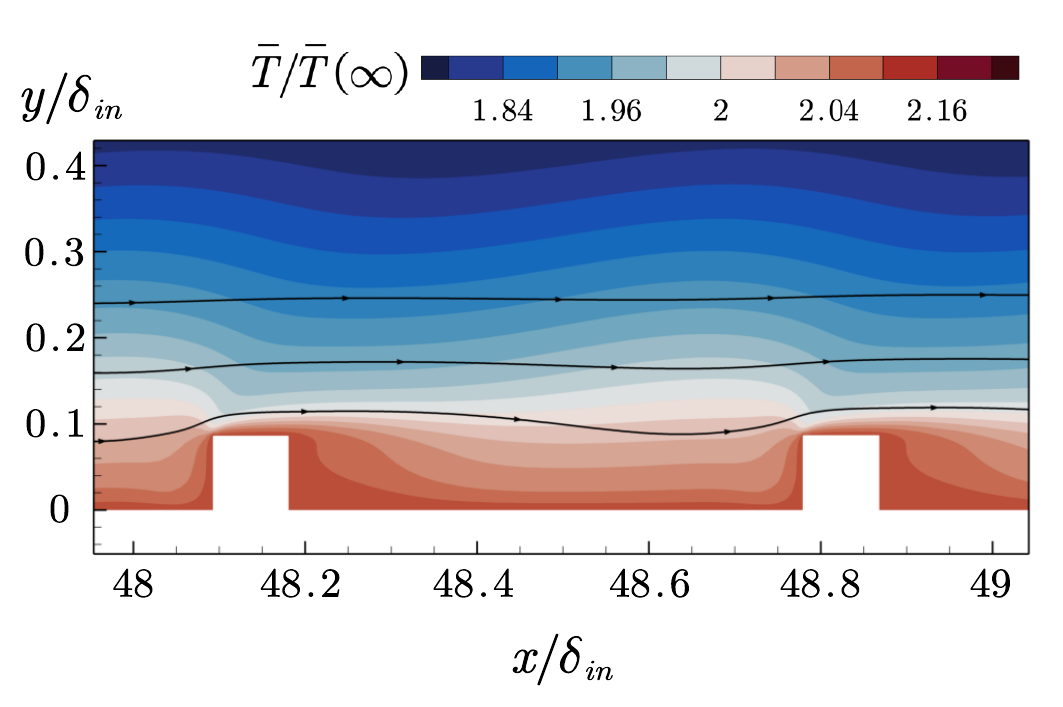}
        \label{fig:T_RA}
    \end{minipage}
    \hfill 
    \begin{minipage}[t]{0.48\textwidth}
        \raggedright 
        (\textit{b}) \\[0.5em] 
        \includegraphics[width=\textwidth]{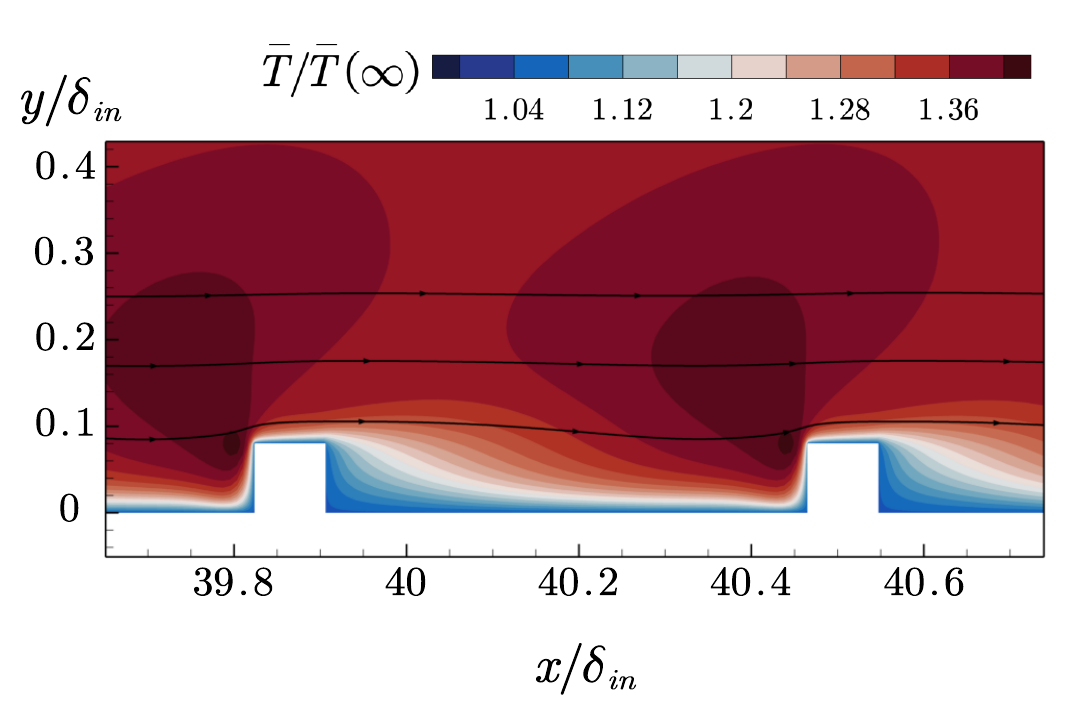} 
        \label{fig:T_RC}
    \end{minipage}
    
    \caption{
        Spatial distributions of the time-averaged temperature: (a) Adiabatic rough-wall case (M2p5RA); (b) cold rough-wall case (M2p5RC). Black lines: streamlines in the vicinity of the roughness heights $k$, $2k$, and $3k$.}
    \label{fig:T_R}
\end{figure}

Conversely, the behaviour in the cold-wall rough case is more complex. Although the rGRA predictions at all three reference heights consistently outperform the classical GRA, the temperature-velocity relation anchored at $k$ still exhibits notable discrepancies with the DNS data in the outer region. This discrepancy stems from the disparate spatial recovery rates of the hydrodynamic and thermal fields. Specifically, while the velocity field recovers to a wall-parallel state relatively quickly above the roughness elements, the thermal field over a two-dimensional rough surface subjected to substantial heat transfer requires a much larger wall-normal distance to achieve spatial homogeneity shown in Figure \ref{fig:T_R}. Consequently, the underlying assumption of Equation (\ref{eq:recovery_rough}) is invalidated at such low elevations. However, as the reference height is elevated to $2k$ and $3k$, thereby bypassing the strong thermal heterogeneity, the rGRA once again provides an accurate description of the temperature-velocity relationship.

\subsection{Strong Reynolds analogy}\label{sec:SRA}

Next, we evaluate the applicability of three variants of the strong Reynolds analogy (SRA) to rough-wall boundary layers. The general formulations of these SRA methods are expressed as
\begin{subequations}
    \label{eq:SRA_formulas}
    \begin{align}
        \text{GSRA:} \quad \frac{T_{\mathrm{r.m.s.}}^{\prime \prime}}{u_{\mathrm{r.m.s.}}^{\prime \prime}} &\approx \Biggl| \frac{\partial \tilde{T}/\partial y}{\partial \tilde{u}/\partial y} \Biggr| \label{eq:GSRA} \\[6pt]
        \text{HSRA:} \quad \frac{T_{\mathrm{r.m.s.}}^{\prime \prime}}{u_{\mathrm{r.m.s.}}^{\prime \prime}} &\approx \Biggl| \frac{\widetilde{v^{\prime \prime}T^{\prime \prime}}}{\widetilde{u^{\prime \prime}v^{\prime \prime}}} \Biggr| = \frac{1}{Pr_t}\Biggl| \frac{\partial \tilde{T}/\partial y}{\partial \tilde{u}/\partial y} \Biggr| \label{eq:HSRA} \\[6pt]
        \text{RSRA:} \quad \frac{T_{\mathrm{r.m.s.}}^{\prime \prime}}{u_{\mathrm{r.m.s.}}^{\prime \prime}} &\approx \frac{C_1}{\sqrt{Pr_t}}\Biggl| \frac{\partial \tilde{T}/\partial y}{\partial \tilde{u}/\partial y} \Biggr|, \quad C_1=1.09 \label{eq:RSRA}
    \end{align}
    \end{subequations}
and the prediction error is defined as
\begin{equation}
    \epsilon = \frac{R}{L} - 1,
    \label{eq:SRA_error}
\end{equation}
where $R$ and $L$ represent the right-hand and left-hand sides of the respective SRA formula.

\begin{figure}
    \centering
    
    \def\myfigheight{4.2cm} 
    
    \def\widthA{0.33\textwidth} 
    \def\widthB{0.292\textwidth} 
    \def\widthC{0.292\textwidth} 

    \begin{minipage}{\widthA}
        {\raggedright (\textit{a})\par}
        \vspace{0.2em}
        \centering
        \includegraphics[height=\myfigheight]{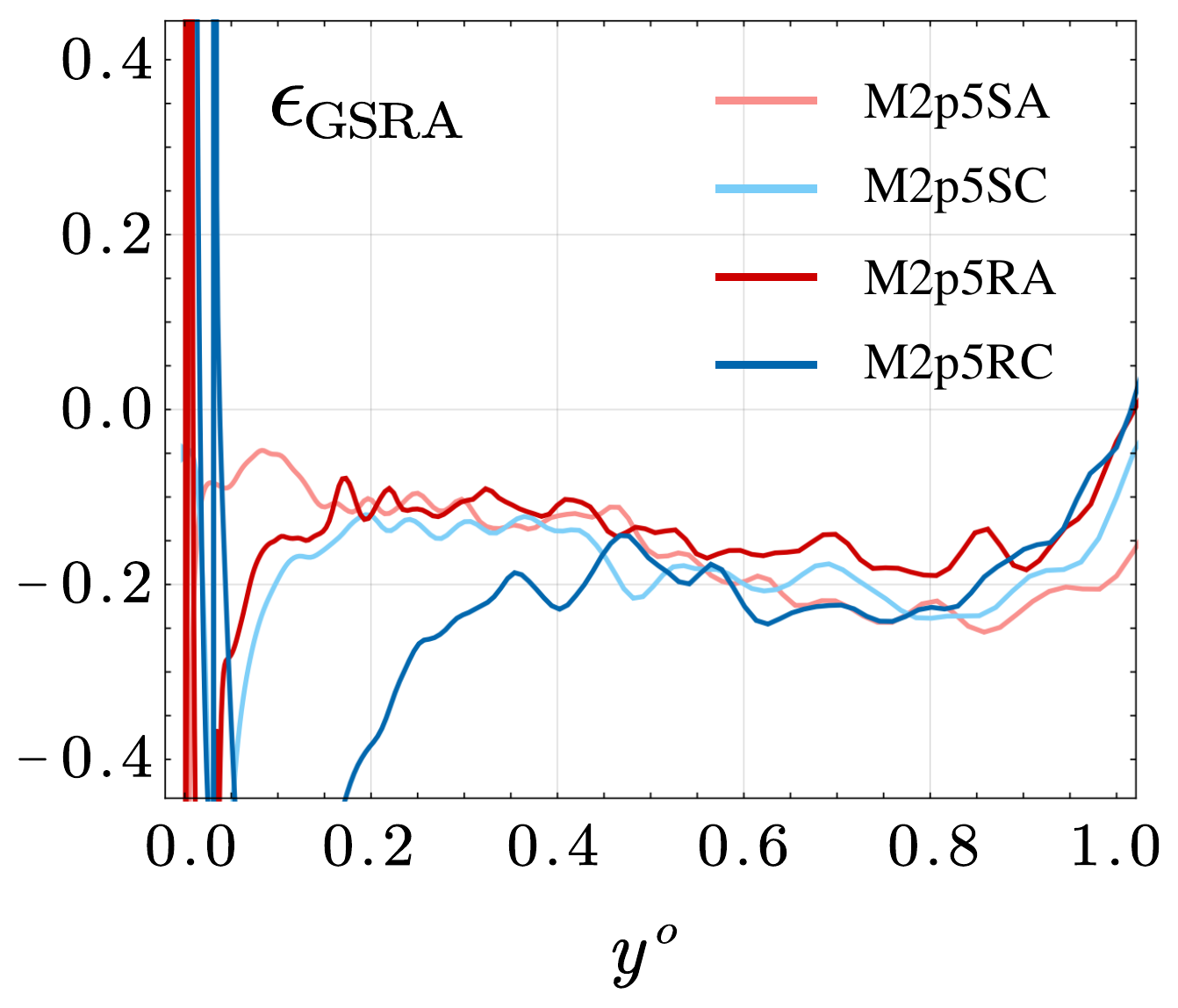}
    \end{minipage}\hfill
    \begin{minipage}{\widthB}
        {\raggedright (\textit{b})\par}
        \vspace{0.2em}
        \centering
        \includegraphics[height=\myfigheight]{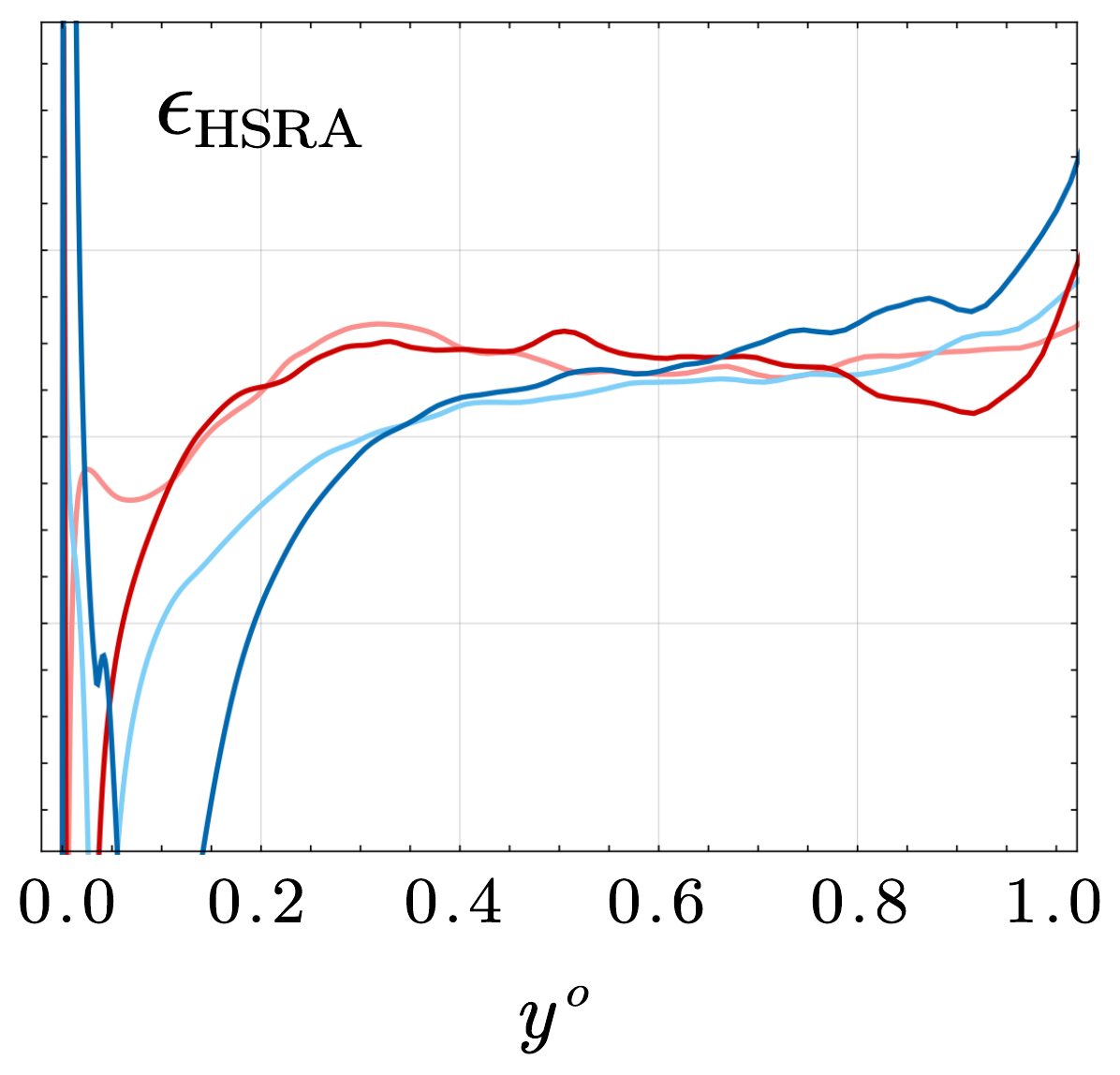}
    \end{minipage}\hfill
    \begin{minipage}{\widthC}
        {\raggedright (\textit{c})\par}
        \vspace{0.2em}
        \centering
        \includegraphics[height=\myfigheight]{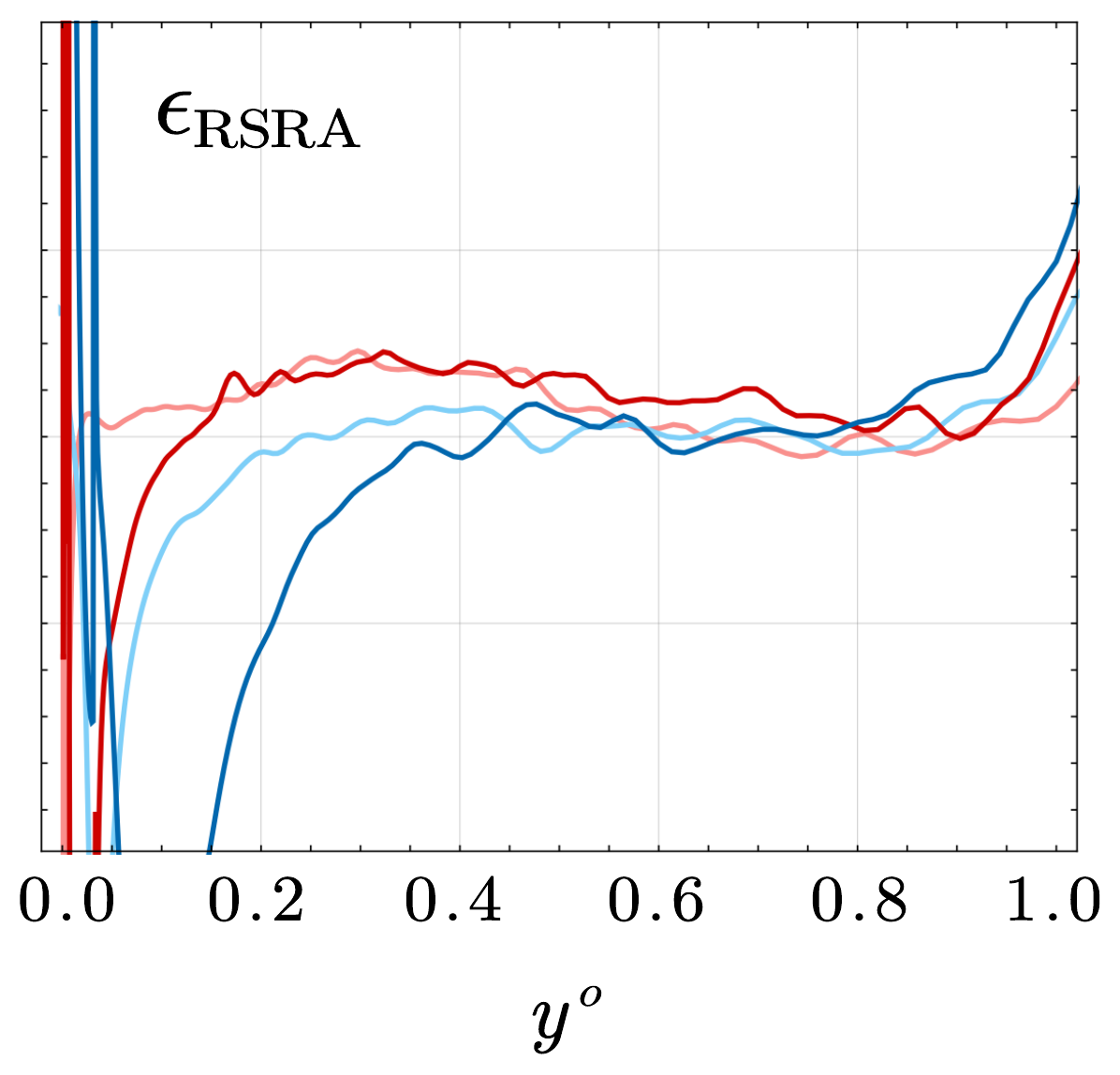}
    \end{minipage}

    \caption{Prediction errors for the three strong Reynolds analogy (SRA) methods in the rough-wall boundary layer. The specific formulations of the SRA methods and the error definition are given in Equations \eqref{eq:SRA_formulas} and \eqref{eq:SRA_error}, respectively. (a) GSRA \citep{gaviglio1987reynolds}; (b) HSRA \citep{huang1995compressible}; (c) RSRA \citep{huang2025refined}.}
    \label{fig:SRA_Error}
\end{figure}

As shown in Figure \ref{fig:SRA_Error}(a), the GSRA proposed by \citet{gaviglio1987reynolds}, which predicts the fluctuation intensities based on the mean wall-normal gradients of velocity and temperature, remains reasonably valid. For both the adiabatic rough-wall (M2p5RA) and smooth-wall cases, the error curves collapse well above $y \approx 0.2\delta$, although the GSRA slightly underpredicts the overall fluctuation levels. For the cold rough-wall case (M2p5RC), however, the GSRA exhibits significant near-wall deviations from the other three cases, and its error profile only converges with the others at a higher elevation of $y \approx 0.5\delta$. This indicates that the combined effect of wall cooling and surface roughness exerts a more profound disruption on the coupling between the fluctuations and the mean flow than on the mean flow itself (as discussed earlier in the context of the GRA), thereby requiring a larger wall-normal distance to recover the analogy.

Turning to the HSRA proposed by \citet{huang1995compressible} (Figure \ref{fig:SRA_Error}b), the error profiles for the rough- and smooth-wall pairs collapse nicely within the region of $0.3\delta < y < 0.5\delta$ under both adiabatic and cold-wall conditions. This suggests that utilizing the Reynolds shear stress for fluctuation prediction still yields satisfactory results, and the wall-normal extent of the roughness influence is marginally lower than that observed for the GSRA. Overall, the HSRA tends to slightly overpredict the fluctuation levels in the outer region. Finally, the RSRA proposed by \citet{huang2025refined} is mathematically the geometric mean of the GSRA and HSRA. Physically, it is grounded in the empirical assumption that the ratio of the generation timescales of the temperature and velocity fluctuating structures remains constant. As depicted in Figure \ref{fig:SRA_Error}(c), the four error curves collapse excellently above $y \approx 0.5\delta$ and approach zero. At lower elevations, however, the RSRA inherits the near-wall sensitivities to both the thermal wall conditions and surface roughness from its GSRA and HSRA counterparts.

\begin{figure}
    \centering
    
    \begin{minipage}{0.48\textwidth}
        {\raggedright (\textit{a})\par}
        \vspace{0.2em}
        \centering
        \includegraphics[width=\textwidth]{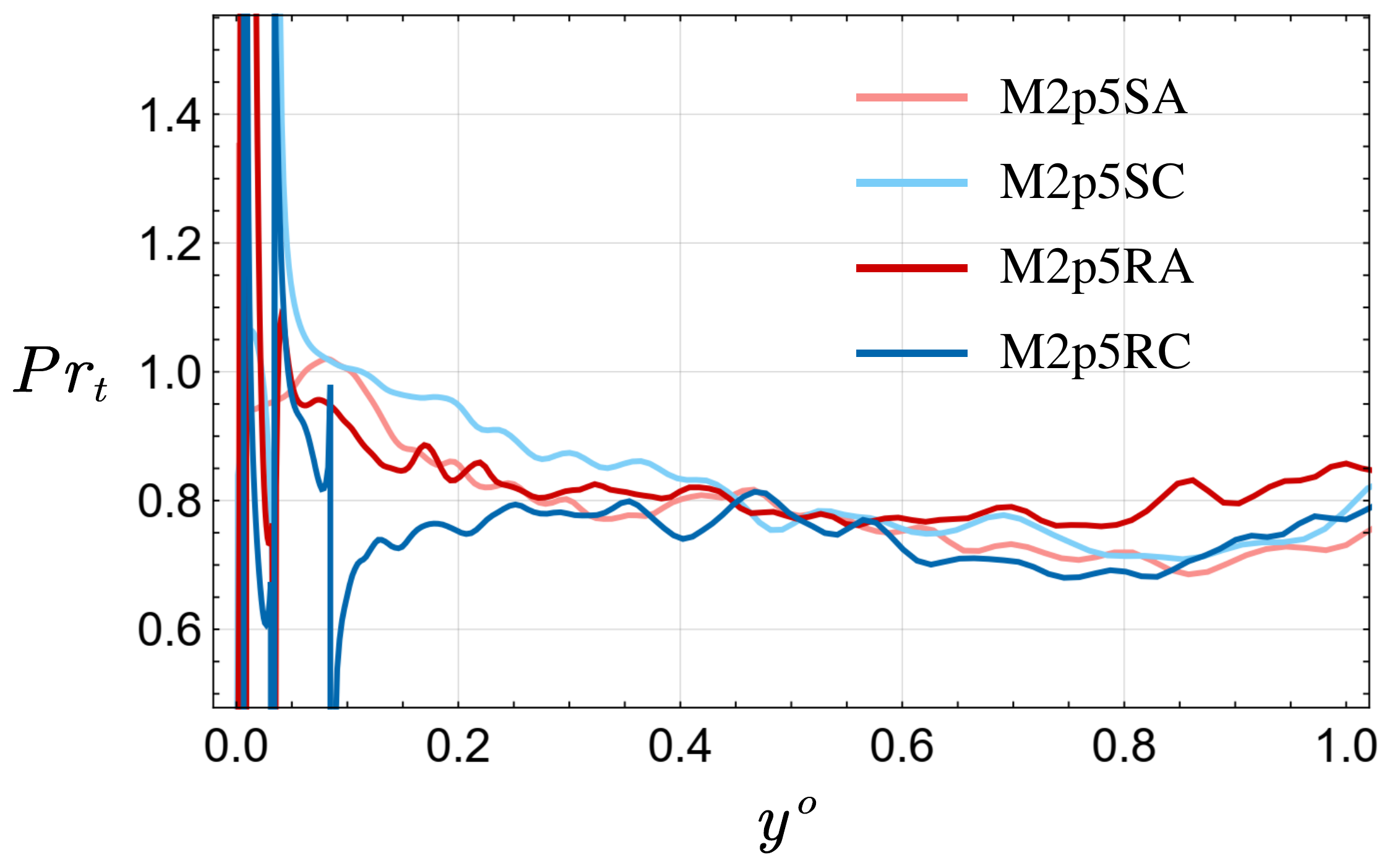}
    \end{minipage}\hfill 
    \begin{minipage}{0.48\textwidth}
        {\raggedright (\textit{b})\par}
        \vspace{0.2em}
        \centering
        \includegraphics[width=\textwidth]{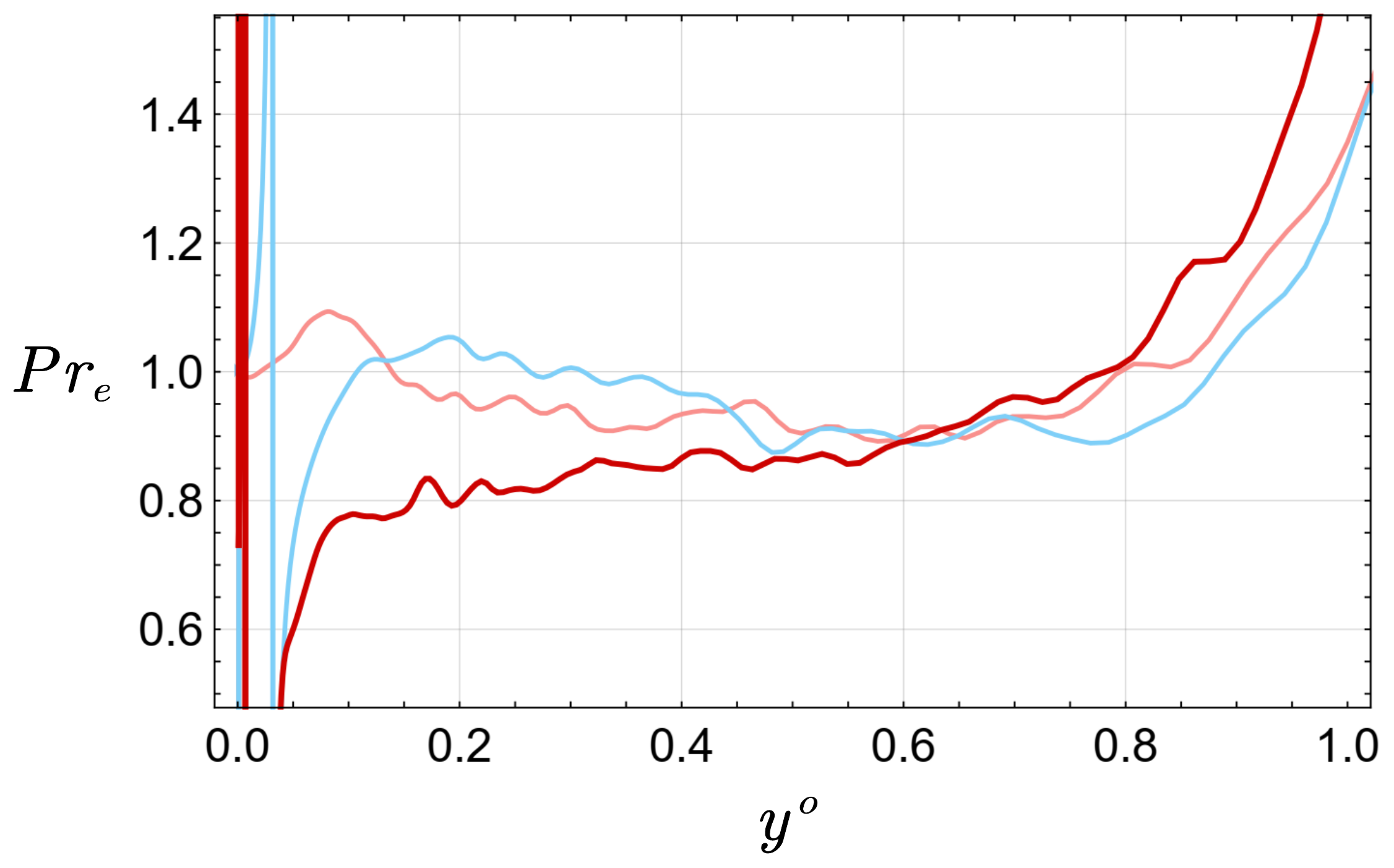}
    \end{minipage}
    
    \vspace{1.0em} 
    
    \begin{minipage}{0.48\textwidth}
        {\raggedright (\textit{c})\par}
        \vspace{0.2em}
        \centering
        \includegraphics[width=\textwidth]{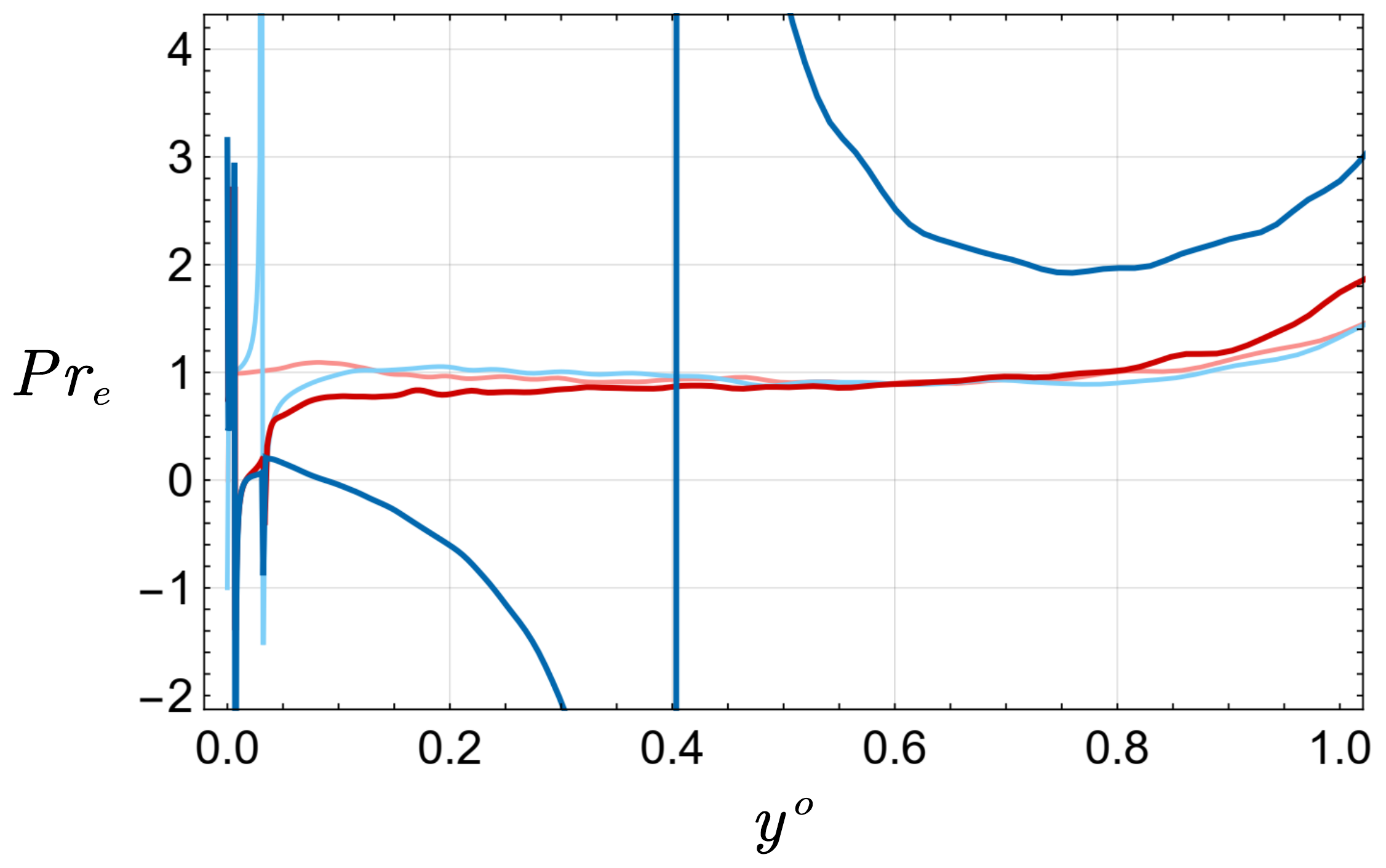}
    \end{minipage}

    \caption{Wall-normal distributions of the Prandtl numbers: (a) Turbulent Prandtl number $Pr_t$; (b) Effective turbulent Prandtl number $Pr_e$ evaluated via the classic GRA; 
    (c) $Pr_e$ profiles for the cold rough-wall case, shown with a wider range to capture the severe deviations.}
    \label{fig:Pr}
\end{figure}

Furthermore, the turbulent Prandtl number ($Pr_t$) employed in the HSRA exhibits a consistent wall-normal distribution. As depicted in Figure \ref{fig:Pr}(a), for both smooth- and rough-wall cases under varying thermal boundary conditions, $Pr_t$ gradually decreases to approximately 0.8 above $y \approx 0.2\delta$. Regarding the effective turbulent Prandtl number ($Pr_e$) defined in the context of the GRA (Equation \ref{eq:diff_vel_temp}), the smooth-wall profiles hover around unity above $y \approx 0.1\delta$ (Figure \ref{fig:Pr}b), gradually decrease to 0.9 near $y \approx 0.4\delta$, and subsequently increase towards the boundary layer edge. This trend is fundamentally consistent with the $Pr_e \approx 1$ assumption inherent in the classical GRA. In contrast, $Pr_e$ for the adiabatic rough-wall case (M2p5RA) is noticeably lower, maintaining a value of roughly 0.8 above $y \approx 0.2\delta$. For the cold rough-wall case, the deviation of $Pr_e$ from unity is even more pronounced (Figure \ref{fig:Pr}c), rendering the $Pr_e \approx 1$ simplification completely invalid.

\begin{figure}
    \centering

    \begin{minipage}{0.48\textwidth}
        {\raggedright (\textit{a})\par}
        \vspace{0.2em}
        \centering
        \includegraphics[width=\textwidth]{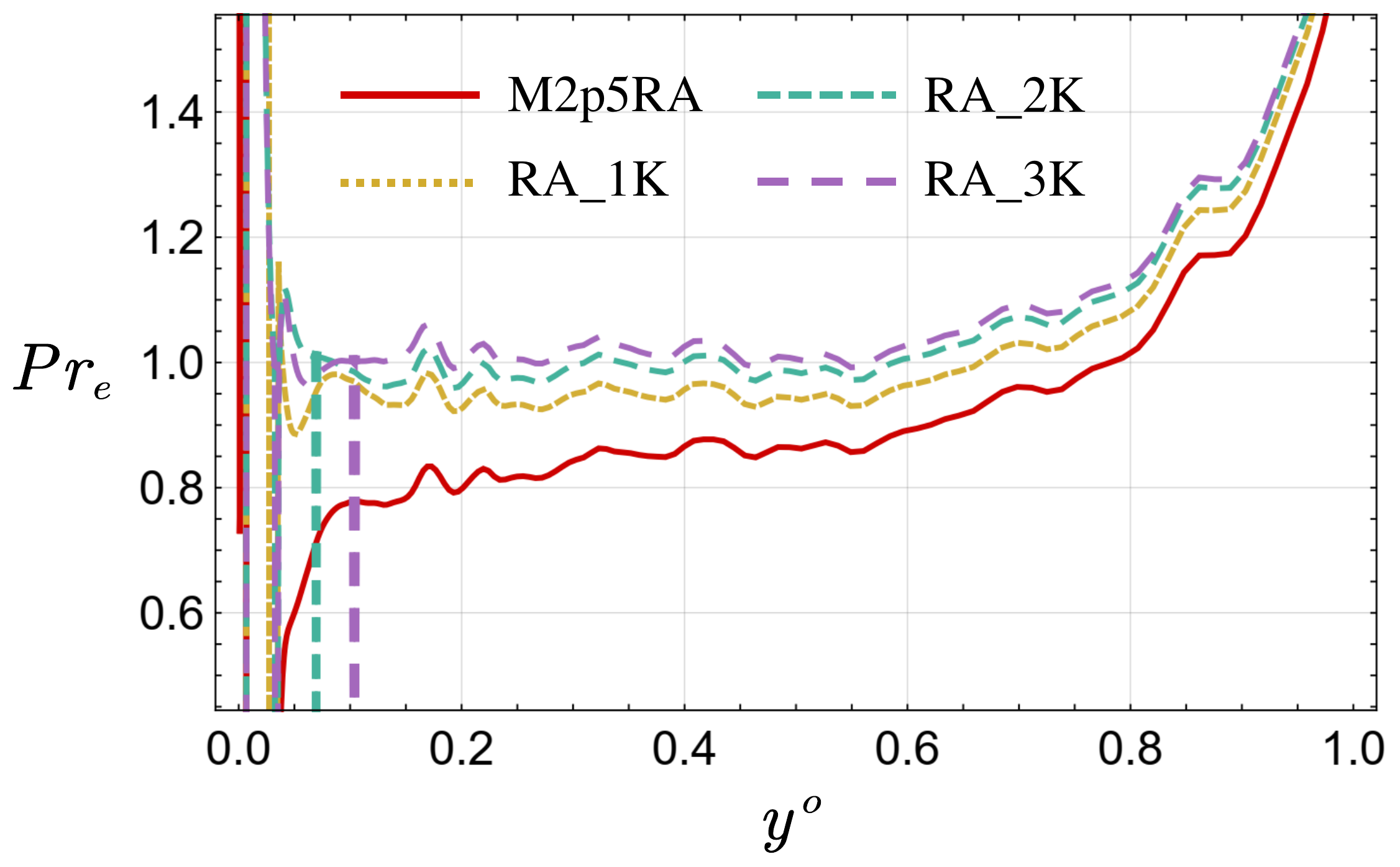}
    \end{minipage}\hfill 
    \begin{minipage}{0.48\textwidth}
        {\raggedright (\textit{b})\par}
        \vspace{0.2em}
        \centering
        \includegraphics[width=\textwidth]{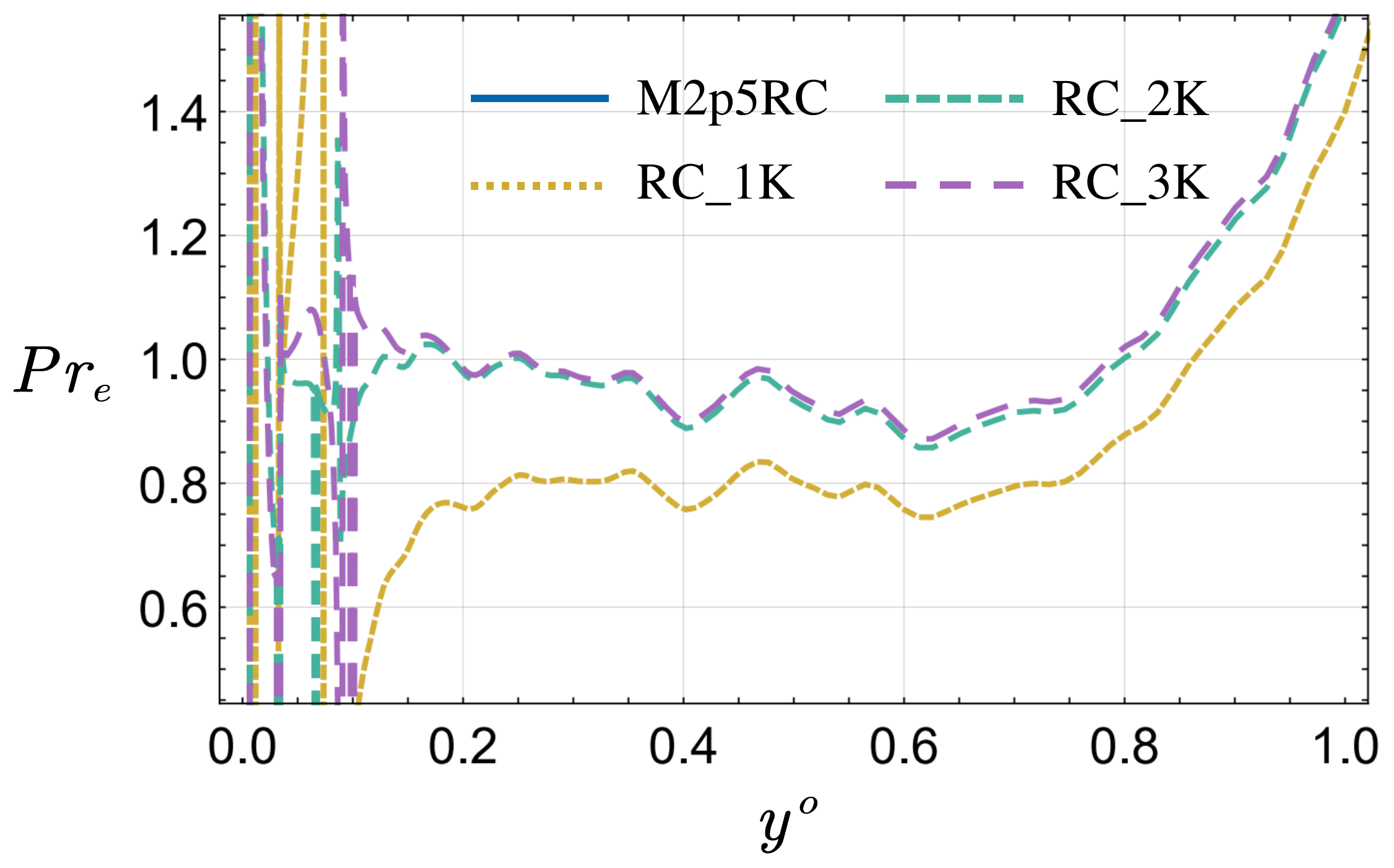}
    \end{minipage}

    \caption{Wall-normal distributions of the effective turbulent Prandtl number $Pr_e$: (a) Adiabatic rough-wall case; (b) cold rough-wall case. Solid lines: GRA; Yellow dashed lines: ${\mathrm{rGRA}}(k,\bar{u})$; Green dashed lines: ${\mathrm{rGRA}}(2k,\bar{u})$; and Purple dashed lines:   ${\mathrm{rGRA}}(3k,\bar{u})$.}
    \label{fig:rGRA_PrE}
\end{figure}

To address this, we apply the rGRA framework formulated in the previous section to re-evaluate $Pr_e$ (Equation \ref{eq:diff_vel_temp_rough}). For the adiabatic rough-wall case (M2p5RA), the $Pr_e$ profiles anchored at reference heights from $k$ to $3k$ progressively converge to unity (Figure \ref{fig:rGRA_PrE}a). For the cold rough-wall case (Figure \ref{fig:rGRA_PrE}b), the classical GRA prediction is omitted from the figure due to its massive deviation from unity. Instead, when employing the rGRA anchored at $k$, the resulting $Pr_e$ value remains near 0.8 for the region below $y \approx 0.8\delta$. As the reference height is further elevated to $2k$ and $3k$, the rGRA predicted $Pr_e$ profiles converge and become invariant. This asymptotic behavior indicates that the selected reference points have reached an elevation characterized by approximately wall-parallel flow. At this stage, the characteristics of the $Pr_e$ profiles align closely with those of the smooth-wall counterparts, thereby further corroborating the accuracy and physical soundness of the proposed rGRA model.

\section{Conclusions}\label{sec:Conclusions}

In the present study, the combined effects of two-dimensional surface roughness and significant wall heat transfer on spatially developing compressible turbulent boundary layers (CTBL) are investigated via direct numerical simulations at a freestream Mach number of $Ma = 2.5$. By employing a high-fidelity immersed boundary method and a streamwise-equilibrious inflow generation strategy, the study systematically evaluates the structural, dynamic, and thermodynamic characteristics of adiabatic and cold rough-wall boundary layers against their smooth-wall counterparts.

Dynamically, we demonstrate that traditional zero-moment methods fail to establish a valid zero-plane displacement that recovers the logarithmic law for the present two-dimensional cavity-type roughness. Instead, a curve-fitting methodology is employed to reproduce the logarithmic region, yielding zero-plane displacements that differ significantly from those obtained via the zero-moment method. Based on these kinematically fitted virtual origins, the performance of velocity transformations is assessed. While the classical van Driest transformation fails to recover outer-layer similarity under non-adiabatic rough-wall conditions, the GFM transformation exhibits remarkable robustness, successfully collapsing the mean velocity defect profiles of both adiabatic and cold rough walls onto the smooth-wall baseline. Furthermore, under the GFM framework, the evaluated roughness function ($\Delta U^+$) for the cold wall remains close to the incompressible reference, whereas the adiabatic wall exhibits a notably higher $\Delta U^+$ compared to both the cold-wall case and its incompressible counterparts.

Thermodynamically, the strong physical asymmetry between momentum form drag and thermal heat conduction significantly disrupts the classical velocity-temperature coupling mechanisms. Consequently, the classical generalized Reynolds analogy (GRA) exhibits substantial discrepancies over rough surfaces, particularly under cold-wall conditions. This failure is fundamentally attributed to the breakdown of the underlying assumption that the effective turbulent Prandtl number remains consistently near unity ($Pr_e \approx 1$) within the roughness sublayer. To overcome this limitation, a rough-wall generalized Reynolds analogy (rGRA) is formulated by introducing an equivalent virtual slip plane or reference-point boundary conditions. By anchoring the rGRA at reference heights sufficiently above the roughness crest ($y \ge 2k$), where the thermal field achieves relative spatial homogeneity, the theoretical temperature-velocity relations are accurately reconstructed.

Finally, evaluations of the strong Reynolds analogy (SRA) models reveal that the coupling between velocity and temperature fluctuations is profoundly sensitive to the combined effects of surface roughness and wall cooling in the near-wall region. Although classical formulations exhibit noticeable near-wall deviations, the refined SRA (RSRA) yields satisfactory predictions of fluctuation intensities in the outer region ($y > 0.5\delta$). Overall, the present findings provide valuable insights into the combined effects of significant wall heat transfer and roughness-induced form drag, offering modified analytical formulations that contribute to the theoretical understanding and modeling of non-adiabatic compressible rough-wall turbulence.

\section{Acknowledgements}\label{sec:Acknowledgements}

This work was supported by National Natural Science Foundation of China under Grant Nos. U2570214, 12425206 and 12388101.

\appendix

\bibliographystyle{jfm}
\bibliography{jfm-instructions}

\end{document}